%% file: agn_compaction.tex
\documentclass[useAMS,usenatbib,a4paper]{mnras}
 \usepackage{epsfig}
 \usepackage{relsize}
 \usepackage{josty}
 \usepackage{amsmath}

\def\dlage{\Delta \log {\rm age}}
\def\dssfr{\Delta \log {\rm sSFR}}
\def\doh{\Delta \log {\rm (O/H)}}

\title[The Morphology-Quiescence Relation] 
{Two Growing Modes and the Morphology-Quiescence Relation in Isolated Galaxies}
\author[Woo \& Ellison] 
    {\parbox{\textwidth}{Joanna Woo$^{1}$\thanks{joannawoo@uvic.ca} \&
    Sara L. Ellison$^{1}$
    } 
\vspace{0.4cm}\\
\parbox{\textwidth}{ 
$^{1}$Department of Physics \& Astronomy, PO Box 1700 STN CSC, Victoria BC V8W 2Y2, Canada\\
}}

\pagerange{\pageref{firstpage}--\pageref{lastpage}} \pubyear{2017}

\hypersetup{draft}
\begin{document}
\label{firstpage}
\maketitle

\begin{abstract}
Quiescence in galaxies correlates strongly with the central density/morphology of the stellar distribution.  We investigate two possible explanations for this morphology-quiescence relation: 1) the central density results from a dissipative core-building event (``compaction'') that feeds an AGN that quenches the galaxy and 2) the central density results from inside-out growth by galaxy-wide star formation that is quenched by processes unrelated to the central density.  We aim to distinguish these two scenarios using the MaNGA survey to determine profiles of stellar age, specific star formation rate (sSFR) and gas phase metallicity (O/H) as a function of stellar mass surface density within 1 kpc ($\sigone$) and total stellar mass ($\Ms$).  We find that gradients in age, sSFR and O/H depend on the galaxy's position on the $\sigone$-$\Ms$ diagram, suggesting at least two evolutionary pathways.  The first pathway consists of galaxies with low $\sigone$ for their $\Ms$ whose centres are old, metal-rich and suppressed in sSFR compared to their outskirts, consistent with the inside-out growth scenario.  The second pathway, consistent with a compaction-like core-building scenario, consists of galaxies with higher $\sigone$ for their $\Ms$, whose centres are younger, enhanced in sSFR and metal-deficient compared to their outskirts.  Moreover, the WISE-selected AGN fraction peaks in the same area of the $\sigone$-$\Ms$ diagram as the core-building pathway.  The sSFR profiles of the quiescent population suggest that galaxies on the compaction-like path quench uniformly, while those on the inside-out growing path quench their centres first.  Our results imply that both pathways contribute to the morphology-quiescence relation.

\end{abstract}


\section{Introduction}
\label{introduction}

It has been known for several decades that galaxies are divided into those that are star forming and those that are quiescent, and yet the physical causes for the cessation of star formation in galaxies (``quenching'') are still a matter of vigorous debate (e.g., \citealp{Bell2008,Peng2010,Peng2012,Woo2013,Omand2014,Woo2015,Zolotov2015,Lilly2016}).  An important observation is the strong correlation between quiescence and bulge-like morphology or high central density.   
Quiescence, as quantified by colour or the star formation rate (SFR), correlates strongly with visual morphological classification (\citealp{Strateva2001,Schawinski2014}), the bulge-to-total ratio or the bulge mass (\citealp{Bluck2014,Teimoorinia2016}), concentration (\citealp{Strateva2001,Kauffmann2004}), the Sersic index (\citealp{Bell2008,Wuyts2011,Vandokkum2011,Mendel2013}), line-of-sight velocity dispersion (\citealp{Vandokkum2011,Teimoorinia2016,Bluck2016}), the surface density within an effective radius (\citealp{Kauffmann2004,Omand2014}), and the surface density within 1 kpc (\citealp{Cheung2012,Fang2013,Woo2015,Woo2017}).  
Furthermore, these correlations are stronger than the correlations between quiescence and other properties such as total stellar mass and the environment \citep{Bluck2014,Woo2015,Teimoorinia2016}.  

Although not all of the above quantities are strictly ``morphology'', they are closely related.  For example, the line-of-sight velocity dispersion is a measure of the random motions typical of stars in a bulge rather than the ordered motions of disks.  The high Sersic index of bulges indicates a strongly peaked density of stars in the centres, or a high central surface density (within 1 kpc or effective radius).  Thus, we refer to the above observations collectively as the ``morphology-quiescence relation''.

Many argue that the morphology-quiescence is a causal relationship between the building of the core/bulge of a galaxy and its eventual quenching (\citealp{Bell2012,Bluck2014}).  Heating by an active galactic nucleus (AGN) has been a particularly compelling hypothesis.  In this picture, dissipative inflows of gas triggered by mergers \citep{Barnes1991,Mihos1996,Hopkins2006,Zolotov2015} or disk instabilities \citep{Friedli1995,Immeli2004,Bournaud2011} feed a central starburst which builds the density of a galaxy's inner core (a process called ``compaction'' in the literature since it results in more compact galaxies - \citealp{Zolotov2015}).  The same inflows feed an AGN which heats and/or expels cold gas from the system preventing further star formation.  

In support of the AGN hypothesis, the directly measured masses of supermassive black holes are tightly correlated with bulge velocity dispersion (\citealp{Kormendy2013}) as well as the specific star formation rate (sSFR; \citealp{Terrazas2016,Terrazas2017}).  AGN hosts also tend to have early-type morphology (\citealp{Sanchez2018}).  Furthermore, the fraction of X-ray detected AGN at high-$z$ peaks for galaxies with high sSFR and high core densities \citep{Kocevski2017}. 
In addition, some tentative evidence that AGN can expel significant amounts of gas from a system has been observed in the local universe (\citealp{Cicone2014,Cheung2016,Baron2018,Bradford2018}) as well as at high-$z$ (\citealp{Nesvadba2008,Nesvadba2017,Bischetti2017}).  (However see also \citealp{Ho2008,Fluetsch2018,Shangguan2018} for a contrary view).  At the very least, winds seem to be ubiquitous in AGN hosts (eg, \citealp{Rupke2017}).

Despite widespread support for the compaction/AGN hypothesis, others, notably \cite{Lilly2016}, argue that the morphology-quiescence relation can be reproduced by a simple model of disk growth that requires no special bulge formation events.  \cite{Lilly2016} propose growing galaxies by successively adding exponential disks of star-forming gas whose scale lengths grow with the expansion of universe.  Disks grown in such an ``inside-out'' manner result in apparently older centres and younger outskirts.  In fact, such inside-out disk growth is expected as a natural consequence of the expansion and turn-around of shells of dark matter in the top-hat spherical collapse model, where the innermost shells, which collapse first, have the lowest angular momentum (\citealp{Kepner1999}; see also \citealp{Fall1980,VandenBosch1998,VandenBosch2002}).  Hydrodynamical simulations (\eg, \citealp{Roskar2008,Tissera2016}) have confirmed that galaxies growing in isolation grow inside-out, producing negative age gradients (older centres, younger outskirts).

\cite{Lilly2016} applied a quenching model to their inside-out growing galaxies such that the quenching probability depended only on its stellar mass $\Ms$, \ie, more massive galaxies are more likely to be quenched.  They did not specify the physical nature of the quenching mechanism, but note that $\Ms$ is a proxy for the halo mass for isolated galaxies which may play a role in quenching (\citealp{Dekel2006,Woo2013,Woo2015}).  
More massive galaxies reach a higher quenching probability earlier than less massive galaxies, and thus they quench with higher density. With their simple model, \cite{Lilly2016} demonstrated that the morphology-quiescence relation can be derived without explicit morphological transformation.  

The processes of secular disk building plus early quenching and dissipative bulge/core-building (``compaction") therefore represent two entirely different mechanisms that have been proposed to explain the observed morphology- quiescence relation in galaxies.  The observation that denser/smaller quiescent galaxies are older than less dense/larger  galaxies \citep{Wu2018} has been interpreted as evidence of the link between galaxy density and the time of quenching \citep{Tacchella2017}.  However, this by itself does not rule out compaction-like scenarios for building old bulges since the more violent forms of disk instability are predicted to occur at high redshift ($z\sim2$), at the peak of the cosmic star-formation history and global accretion rate history \citep{Keres2009,Zolotov2015}. 

Distinguishing between core-building compaction-like scenarios and progenitor effects of inside-out growth depends crucially on the presence or lack of new star formation in the centres of galaxies.  If galaxies are growing by secular inside-out disk growth by default, then the newest stars are found in the disk.  A compaction event on the other hand, adds new stars to the centres of otherwise inside-out growing disks.  Simulations predict that profiles of SFR density rise in the centres during compaction events \citep{Tacchella2016}.  
Thus, if we assume that the effect of subsequent mergers is minimal (more on this in \secref{caveats}), the signature of secular disk growth is a negative age gradient whereas a compaction-like event will flatten the age gradient.

With the recent advent of large integral field unit (IFU) surveys of nearby galaxies such as Mapping Nearby Galaxies at the Apache Point Observatory (MaNGA - \citealp{Bundy2015}), Calar Alto Legacy Integral Field spectroscopy Area (CALIFA - \citealp{Sanchez2012}) and Sydney-AAO Multi-object Integral field spectrograph survey (SAMI - \citealp{Croom2012}),
we can begin to study the age gradients of galaxies as a function of global properties, which is the subject of this paper.  
Indeed, some papers have already presented stellar population gradients in CALIFA disks (\citealp{Sanchez-Blazquez2014}) and in the general population in MaNGA (\citealp{Goddard2017,Wang2017}), finding various trends of the age gradient with global properties.  However, these studies often disagree as to the magnitude and even direction of the population gradients.  Here we aim to extract the properties of the gradients that are robust using a novel method for searching for gradients in a differential manner.  We also explore what these gradients mean in the context of our current understanding of structural growth and quenching.

Our goal is to explore whether compaction-like core-building events contribute to the build-up of the central density of galaxies and whether this is related to quenching.  Thus, this paper is a study of the gradients and profiles of stellar age, sSFR and gas phase metallicity as a function of the central density and $\Ms$.

\section{Data}
\label{data}

\subsection{Sample}
\label{sample}

Our sample is drawn from the MaNGA survey (\citealp{Bundy2015}).  At the time of analysis, the most recent public release was the Data Release 14 (DR14), containing 2780 galaxies.
We downloaded the publicly available data cubes\footnote{\texttt{https://www.sdss.org/dr14/manga/manga-data/data-access/}} and the summary table for the MaNGA Data Reduction Pipeline\footnote{\texttt{https://data.sdss.org/sas/dr14/manga/spectro/redux/v2\_ 1\_2/drpall-v2\_1\_2.fits}}.
The MaNGA targets were drawn from the NASA-Sloan Atlas (NSA) catalogue (\citealp{Blanton2011}).  In order to obtain global properties for each MaNGA object, the targets were matched to the Sloan Digital Sky Survey (SDSS) DR7 spectroscopic catalogue, requiring a match separation of $<3''$.  This produced 2674 matches with a unique SDSS \texttt{objID}.

We draw upon various public catalogues of global properties with inhomogeneous sample definitions, including catalogues of group membership (\citealp{Yang2012} - 2392 matches), global star formation rate (\citealp{Brinchmann2004} - 2447 matches), disc inclination measurements (\citealp{Simard2011} - 2541 matches), and products from the DR7 photometric pipeline (2226 matches; see the following subsections for details).  
We have found that 2148 MaNGA objects appear in every catalogue that we use.  

Since we are interested in the compactness of galaxies as measured by the density within 1 kpc, we restrict our sample to galaxies with $z < 0.07$, as well as those with point spread function (PSF) width in the $r$-band less than 1 kpc in radius.  This reduced the sample to 1790 objects.  Next, we removed galaxies for which the disc inclination (from the catalogue of \citealp{Simard2011}) $b/a < 0.4$ resulting in 1262 galaxies.  This cut was meant to minimize significant errors due to dust.   We also required the total masses to be $\Ms > 10^{9}\Msun$, yielding 1184 galaxies.  

The questions posed in this study concern the quenching and morphological evolution of {\it isolated} galaxies since galaxies in groups and clusters are expected to experience additional quenching mechanisms due to their environment (\citealp{Gunn1972,Larson1980,Read2006,Villalobos2012,Woo2015,Woo2017}).  The morphological evolution of group/cluster galaxies will be addressed in future work.  In our present study, we restrict our sample to those galaxies that belong in Yang et al. groups with only 1 member.  We call these ``isolated'' galaxies.  This reduces the sample to 616 galaxies.  

Lastly, we applied various quality control cuts to the age, sSFR and gas phase metallicity measurements in the MaNGA cubes (see below) which reduced the sample to 
610 galaxies for studying stellar properties (namely age), and 482 galaxies when studying gas properties (sSFR and O/H).  
In all the following calculations we have assumed the following cosmology: $\Omega_{\rm m} = 0.3$, $\Omega_{\Lambda}=0.7$ and $h=0.7$.

\subsection{Global Quantities}
\label{global}

In order to facilitate easy comparison with other studies, we use the stellar masses ($\Ms$) and star formation rates (SFR) from the widely used MPA-JHU DR7 catalogues\footnote{\label{mpa}\texttt{https://wwwmpa.mpa-garching.mpg.de/SDSS/DR7/index.html}}  (\citealp{Kauffmann2003a,Brinchmann2004,Salim2007}), which adopt the Kroupa IMF.  We define ``star-forming'' (SF) and ``quiescent'' (Q) galaxies as being above or below the commonly used division of log SFR/$\Ms$ (yr$^{-1}$) $= -11$.

In order to parametrize the central density, we compute the stellar surface density within 1 kpc ($\sigone$).  The primary reason for this is that $\sigone$ is a measure of what happens to the core (inner kpc), whereas a scale radius (e.g. a half-light or half-mass radius) also responds to events happening in the outskirts.  Effective quantities are prone to larger fluctuations.  Simulations have shown that even a steep decrease in the effective radius during a compaction event can be followed by a significant increase due to the re-accretion of a disk and/or interactions (\citealp{Wellons2016,Barro2017,Zolotov2015}).  $\sigone$, however, is much more stable to disk regrowth and interactions (\citealp{Wellons2016,Barro2017}).  It also more tightly correlates with $\Ms$ with much slower evolution with redshift (\citealp{Barro2017}).  
The actual scale of 1 kpc (as opposed to 2 or 0.5 kpc) was chosen as a seeing-constrained measure of the core in the nearby universe.  However, 1 kpc also happens to corresponds to the radius at which pseudobulges begin to build the inner density (\eg, \citealp{Fisher2010}).  

We computed $\sigone$ following the method of \cite{Fang2013}.  Using the $\Ms$ estimates from the MPA-JHU catalogue, \cite{Fang2013} computed the relation between stellar mass-to-light ratio in the $i$-band ($\Ms/L_i$) and rest-frame $g-i$ colour.  The rest-frame $g-i$ colour was computed from total \texttt{model} magnitudes from SDSS DR7, which were corrected for Galactic extinction and k-corrected to $z=0$ using the code of \cite{Blanton2007}.   Following the same method, we derived the relation between $\Ms/L_i$ and $g-i$ colour to be $\log \Ms/L_i = 1.13 + 0.83(g-i)$ in AB-centric units with a scatter of 0.04 dex, which is the same as the relation in \cite{Fang2013} (within the bootstrap errors of $\sim 0.1$ in each parameter).

We retrieved the $ugriz$ surface brightness profiles from the DR7 Catalogue Archive Server (see \citealp{Woo2015} for details).  The profiles were corrected for Galactic extinction using the extinction tags in the \texttt{SpecPhoto} table (which are derived from the dust maps of \citealp{Schlegel1998}), and were k-corrected to $z=0$ using the code of \cite{Blanton2007}.  We estimated a stellar mass for each annulus using the relation between $\Ms/L_i$ and $g-i$ colour derived above.
In this way, we were able to 
construct stellar mass profiles as well as their cumulative profiles.  We then interpolated the cumulative profiles to find the mass within 1 kpc.  
In the same way, we interpolated the cumulative mass profiles to find the radius that contains half the total mass ($R_{\rm e,mass}$) from the MPA-JHU catalogue. 

\subsection{MaNGA Stellar Ages}
\label{agedef}

For each data cube from the MaNGA data release, we computed the signal-to-noise ratio in each spaxel ($\sntot$) averaged over the observed wavelength range that corresponded within the rest-frame range of 5590-5680 \AA, which is a range that is relatively devoid of emission lines and sky lines.
We used the Voronoi binning code of \cite{Cappellari2003a} to adaptively bin adjacent spaxels such that their resulting $\sntot$ reaches our target value (determined below).  Before performing this binning, we masked out foreground stars.  We also masked spaxels with low $\sntot$ ($<2$), as recommended in the documentation of the code, which improved the contiguity of the binning maps of some objects with many low-$\sntot$ spaxels. 
We refer to the resulting Voronoi bins of spaxels as ``baxels''.

To account for the covariance of the noise between adjacent spaxels, we multiply the quadrature-added noise of the binned spaxels by the correction factor computed by \cite{Law2016}: $1.0 + 1.62\log(n)$, where $n$ is the number of binned spaxels, when $n< 100$.  For $n$ $\gteq$ 100, the correction factor is a constant 4.2.  
For some galaxies, the total S/N of the entire cube is such that a target $\sntot$ of 
20 per baxel cannot be reached and the algorithm fails.  Thus instead of a fixed target $\sntot$, we use either 
20 or the maximum $\sntot$ of all spaxels, whichever is lower.  In the latter case, this results in a binning map where one baxel contains only one spaxel while the rest of the baxels consist of several spaxels.  We save the target S/N as well as the actual $\sntot$ of each baxel in order to use it in our error analysis below.  
Only about 2\% of galaxies were binned to a target S/N less than 10, and 8\% of galaxies were binned to a target S/N less than 15. 
Applying this binning to the entire MaNGA DR14 sample of 2780 galaxies resulted in 
716064 baxels.  
However in the ensuing analysis, we use only those objects that pass the cuts described above in \secref{sample}.  

We performed full spectral fitting on each baxel using the Penalized Pixel Fitting (pPXF) algorithm of \cite{Cappellari2017} (v6.7.13\footnote{We made one small modification to this version to remove the numerical limits on the coefficients of the multiplicative polynomials.  This was to allow for cases of extreme reddening (M. Cappellari private communication).  We have confirmed that the polynomials are always $> 0$ for all baxels.}, which is an upgrade of the original algorithm presented in \citealp{Cappellari2004}).  Before fitting, each baxel is corrected for foreground Galactic extinction using the attenuation curve of \cite{Fitzpatrick1999}.  
The SSP models that are fitted to the baxels are the E-MILES templates \citep{Vazdekis2012,Ricciardelli2012,Vazdekis2016} using the \cite{Kroupa2001} IMF and BaSTI theoretical isochrones (\citealp{Pietrinferni2004}).  Since our fitting wavelength range includes 7385-8950 \AA~ which covers the CaII triplet, we limit the grid of SSP models to those in the ``SAFE'' ranges\footnote{\texttt{http://www.iac.es/proyecto/miles/pages/ssp-models/ safe-ranges.php}} of metallicity (6 grid points $-0.66 \lteq$ [M/H] $\lteq 0.26$) and age (52 grid points for 30 Myr to 13.5 Gyr) for using this wavelength range.  Since our final sample consists of low-redshift ($z < 0.07$) massive ($\Ms > 10^{9}\Msun$) objects (see \secref{sample}) we do not expect metallicities to fall outside of this safe range\footnote{We also performed the analysis using the more standard range of 3500-7429.4 \AA, included metallicities down to [M/H] = -1.2, and found that our results do not change significantly.}.  This produces an age-metallicity grid of 312 templates.  We note that we performed the same analysis using the templates of \cite{Maraston2011} and our results are qualitatively similar.  

The pPXF algorithm computes the weights that produce the best-fitting linear combination of the templates that reproduces the observed spectrum.  It fits emission lines simultaneously such that the stellar population can be fitted despite emission filling in important absorption lines.   We have included the Balmer lines down to H$\delta$, as well as HeI, [SIII], [ArIII], and the doublets of [OII], [SII], [OIII], [OI] and [NII].  All forbidden lines are fixed to share a common, single, velocity component.  The Balmer lines are also fixed to a common velocity, but the Balmer velocity component is allowed to be distinct from the velocity component of the forbidden lines. 
We use the weights computed by pPXF to estimate the mean mass-weighted log age, metallicities and $i$-band mass-to-light ratio, log $\Ms/L_i$, for each baxel.

We tested the accuracy of the mass-weighted stellar ages and metallicities computed by pPXF by fitting simulated spectra with known ages and metallicities, and with the same wavelength range, resolution and noise typical of our baxels.  We found that in the majority of cases, both the ages and metallicities are recovered with less than 0.2 dex scatter.  We found very minimal offsets of less than 0.1 dex.  Full details of our tests are presented in Appendix A.

From our tests, baxels had ``good'' ages if they fulfilled the following criteria: 
\begin{enumerate}
\item the $\chi^2$ of the fit $< 3$ (96\% of baxels)
\item Mass-weighted log Age/yr $>$ 8.5 ($> 99$\% of baxels);
\item Light-weighted log Age/yr $>$ 8.0 ($99$\% of baxels);
\item The difference between the mass-weighted and light-weighted age, log Age$_{\rm MW}$ - log Age$_{\rm LW}$ $<$ 1.1 (97\% of baxels);
\item Mass-weighted [M/H] $<$ 0.255 (99\% of our baxels); this cut removes the measurements that are saturated at the very highest metallicities (0.26);
\item log $\Ms/L_i$ $>$ -0.3 (95\% of our baxels).
\item $\sntot > 10$ (95\% of baxels)
\item the baxel contains $< 10$ spaxels (92\% of our baxels)
\item the baxel has $> 2500$ valid wavelength pixels (non NaNs) ($> 99$\% of our baxels).
\end{enumerate}
The first seven criteria were motivated by our tests in Appendix A.  In total, 
77\% of the baxels in 
97\% of the galaxies passed all these cuts.  However, we have found that our results are not sensitive to the actual values of our cuts within reasonable ranges.  With these cuts, our tests predict a 1-$\sigma$ error on the mass-weighted mean log Age of 0.17 dex and in log $\Ms/L_i$ of 0.11 dex.
After matching with the SDSS DR7 catalogues and applying the cuts in \secref{sample}, we are left with 
107966 baxels in 610 galaxies. This is the sample of baxels used when studying stellar age gradients and profiles.  

Note that pPXF includes the option to regularize the solution such that the weights represent a more physical, smooth star formation history and metal enrichment evolution.  Our tests have indicated that the accuracy of the mean mass-weighted log age and metallicities for the regularized solutions is similar to the non-regularized version, but the computation time is more than 10 times longer when regularized.  Therefore we have opted to compute the non-regularized solutions for this study.

\subsection{Surface Densities of Stellar Mass, Star Formation Rates and Gas Metallicities}
\label{sfdef}

The stellar mass surface density ($\Sigma_{*}$) in each baxel is computed from the $i$-band luminosity density of the observed spectrum and the $\MLi$ obtained from the pPXF weights (assuming the \citealp{Kroupa2001} IMF). 
pPXF simultaneously fits the emission line flux densities together with the stellar populations. 
We use these flux measurements to compute SFR and gas metallicity for the baxels.  The Voronoi binning allows for more accurate subtraction of the stellar contribution, and accounts for the correlated errors between adjacent spaxels when computing S/N for the emission lines.  

We computed the gas reddening from the Balmer decrement and dust-corrected the flux densities for all lines using the \cite{Cardelli1989} extinction curve.  We then computed the SFR density from the corrected $\Ha$ luminosity densities using the expression given in \cite{Kennicutt2009} (divided by the baxel size): 
 $\Sigma_{\rm SFR}(\Msun {\rm yr}^{-1} {\rm kpc}^{-2}) = 5.5 \times 10^{-42} \Sigma_{L,{H\alpha}}({\rm erg~ s}^{-1}{\rm kpc}^{-2})$, 
which is valid for electron temperatures of $10^4$ K and Case B recombination, and assumes the \cite{Kroupa2001} IMF.  The specific star formation rate for each baxel (sSFR) is then $\Sigma_{\rm SFR}/\Sigma_{*}$.  

Gas-phase metallicities are computed using the \cite{Marino2013} calibration: 12 + log(O/H) = $8.533 - 0.214(y-x)$, where $y = \log \OIII / \Hb$ and $x = \log \NII / \Ha$. 

When studying the profiles of sSFR and O/H, we applied the following cuts to the baxels:
\begin{enumerate}
 \item cuts (viii) and (ix) listed in \secref{agedef}; the emission line fitting tends to be robust against errors in the stellar population fitting, so we do not apply the other criteria;
 \item the criterion for ionization from star-formation (no composites): $y < 0.61/(x- 0.05) + 1.3$ (\citealp{Kewley2006}), where $y = \log \OIII / \Hb$ and $x = \log \NII / \Ha$;
 \item the S/N $>$ 3 for the four emission lines used in (ii);
 \item E(B-V) $<$ 1.3 from the Balmer decrement (only one galaxy had baxels with extremely high E(B-V)).
 \end{enumerate}
According to these criteria, 
38\% of the baxels inside 66\% of the galaxies  contain gas ionised primarily by star-formation with sufficient S/N.
This corresponds to  
76065 baxels in 482 galaxies that also meet the criteria in \secref{sample}.
This is the sample of baxels used in the analysis of gas profiles and gradients (\ie, of sSFR and gas metallicity).

\subsection{Gradients and Relative Offsets in Age, sSFR and O/H}
\label{graddelta}

Gradients of stellar age, sSFR and gas metallicity for each galaxy were computed as the slope of the linear fit of the quantity of interest vs. the galactocentric distance of the baxels (from the centres measured in the NSA catalogue - \citealp{Blanton2011}) in units of the half-mass radius $\Remass$.  
A gradient was only computed if there were at least 10 good baxels (as determined by the criteria in \secref{agedef} and \ref{sfdef}).  
15 objects did not fulfill this criterion when computing the age gradients, while 58 objects had fewer than 10 good baxels for measuring
sSFR and O/H gradients.
We also downloaded the public age gradients of \cite{Goddard2017} which are computed using the Firefly code \citep{Wilkinson2017}.  There are some differences in these gradients, but certain features are robust which we describe in \secref{agegradients_sf} and \ref{caveats}.

\begin{figure}
\includegraphics[width=\linewidth]{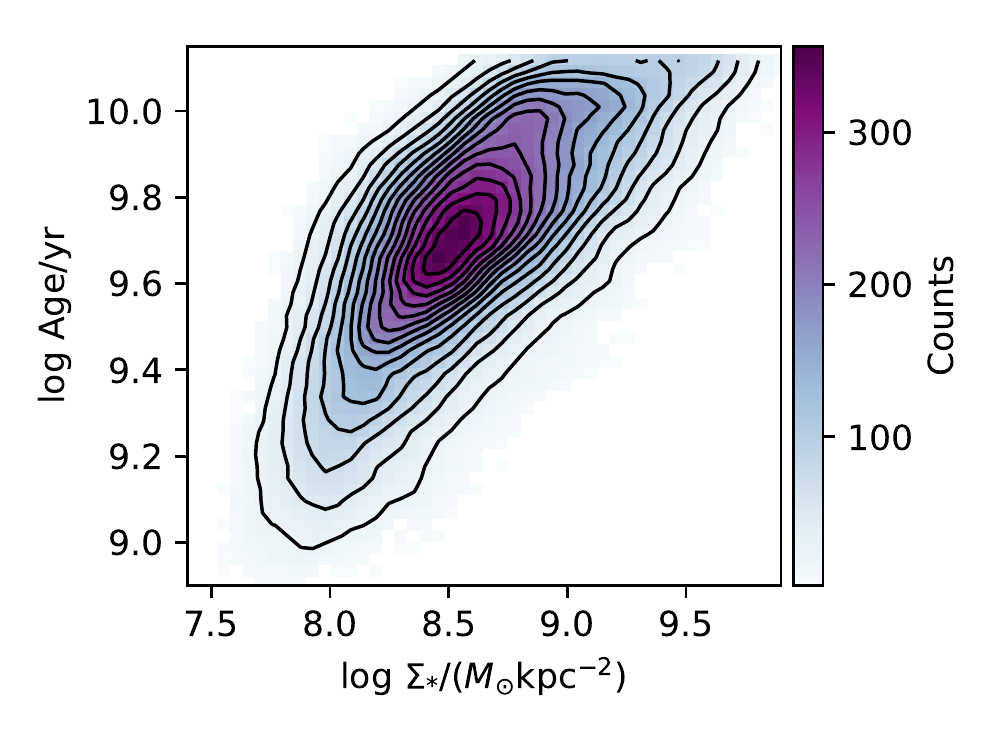}
\caption{\small Mass-weighted stellar age vs. stellar mass density in MaNGA baxels.} 
\label{agevssigs}
\end{figure}

One of our goals is to study profiles of relative {\it enhancements} and {\it deficits} in stellar age, sSFR and O/H ($\dlage$, $\dssfr$, and $\doh$) between galaxies with different stellar mass density profiles.  Thus, we implemented a novel method to control for other factors that might correlate with these three quantities.  For example,  \fig{agevssigs} shows that stellar age correlates strongly with $\sigs$.  We can remove this correlation by defining $\dlage$ as the difference in log age of a baxel from the mean log age of all baxels within a 0.1 dex window of the test baxel's $\sigs$.  In addition to $\sigs$, we also control for total stellar mass ($\Ms$) and radial position ($R$) within the galaxies using windows of 0.1 in log $\Ms/\Msun$ and 0.2 in $R$/kpc.  
These windows in $\sigs$, $\Ms$ and $R$ are iteratively grown by 0.05, 0.05 and 0.1 if the number of control baxels is less than 10, but in practice, this happens for $< 1\%$ of the baxels.  A typical galaxy of any mass will have $\dlage = 0$ at all $R$ by definition.  $\dssfr$ and $\doh$ are computed in the same way.  Note that $\Delta \log {\rm SFR}$ will be the same as $\dssfr$ due to our controlling for $\sigs$.  In the computation of $\dlage$, the sample of ``good'' age baxels are used for the pool of controls as defined in \secref{agedef}, while for $\dssfr$ and $\doh$, the pool of controls are the ``good'' gas baxels defined in \secref{sfdef}.

We then computed profiles of $\dlage$, $\dssfr$, and $\doh$ for different populations, and these are shown in the ``b'' panels of \mfigs{ridgeline_age}{ridgeline_gasmet_q} (discussed below).   These profiles are the median of all baxels within radial bins and then smoothed with a box filter kernel of 3 radial bins.  The bin ``centres'' are the median $R$ of all baxels in the bin.  Radial bins are discarded from the plotted profiles if the number of baxels in the bin is $< 2\%$ of the total number of baxels, or if the number of galaxies represented in a radial bin is $< 10\%$ of the total number of galaxies in the population.  Errors on the profiles are the standard error on the median.  

``Gradients'' and ``profiles'' provide complementary information regarding the structure of galaxies.  Gradients are a single number per galaxy in age, sSFR and O/H, and are the measurements presented in the left panels of \mfigs{ridgeline_age}{ridgeline_gasmet_q} (discussed below).  The profiles of our ``$\Delta$'' quantities are computed as a median for a population of galaxies, and captures more radial information.  Profiles are the measurements presented in the right panels of \mfigs{ridgeline_age}{ridgeline_gasmet_q} (discussed below).

\section{The Growth of Galaxies on the $\sigone$-$\Ms$ Relation}
\label{sfridgelinesec}

\subsection{The $\sigone$-$\Ms$ as a Map for Galaxy Growth}

\begin{figure}
\includegraphics[width=\linewidth]{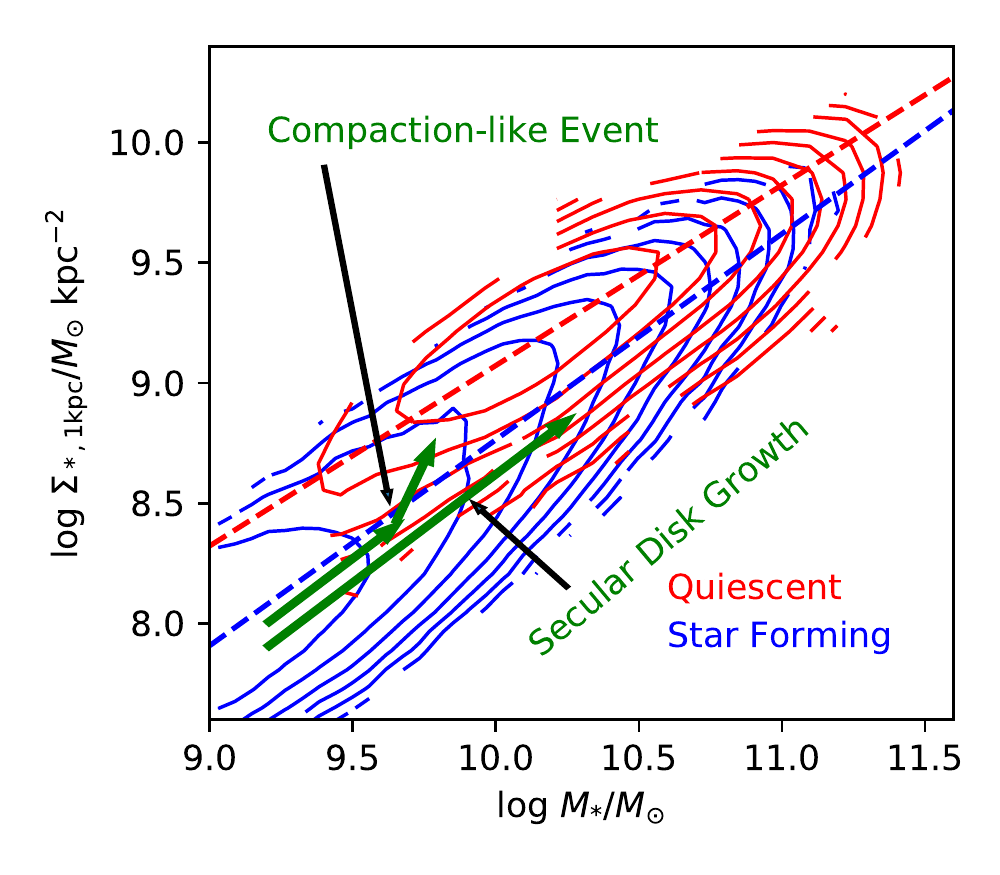}
\caption[Caption]{\small The surface density within 1 kpc as a function of total stellar mass for isolated quiescent galaxies (red) and star forming galaxies (blue) in the SDSS DR7 within 0.01 $< z < $ 0.07.  The contours are weighted by the 1/$V_{\rm max}$ multiplied by the inverse of the spectroscopic completeness.  
The dashed blue and red lines are the volume-weighted least-squares fits to the SF and Q populations 
($\log \sigone = 0.86 (\log \Ms/\Msun -10.25) + 8.98$ and $\log \sigone = 0.75 (\log \Ms/\Msun -10.25) + 9.26$)\footnotemark.
The green arrows qualitatively indicate the expected evolutionary paths of galaxies growing in the two modes indicated.  The secular disk growth track increases in $\Ms$ faster than in $\sigone$, so the slope of this growth should be $\ltsima$ 1 (the slope of the arrow is about 0.8).  The compaction track may start off on the secular track, but when a galaxy experiences a compaction event this leads to an increase in $\sigone$ that is faster than in $\Ms$, and brings a galaxy to the upper part of the diagram. }
\label{ridgeline}
\end{figure}

Our goal as stated in the Introduction is to determine the stellar age gradients of galaxies as a function of position on the $\sigone$-$\Ms$ relation.  The distribution of galaxies in the $\sigone$-$\Ms$ plane was first studied by \cite{Cheung2012} and \cite{Fang2013} who noted that quiescent galaxies lie on a tight sequence in this parameter space.  We reproduce this relation for reference in \fig{ridgeline} (red contours) for the quiescent (Q) galaxies in the DR7 SDSS (defined as having global sSFR $< 10^{-11} {\rm yr}^{-1}$, where the global sSFR is from the \citealp{Brinchmann2004} DR7 catalogue, about 15000 galaxies).  We measure a scatter of 0.17 dex for these quiescent galaxies.
Star-forming (SF) galaxies (blue contours, defined as having global sSFR $> 10^{-11} {\rm yr}^{-1}$, about 41000 galaxies) at the same mass have a wider range of $\sigone$ than quiescent galaxies, but their $\sigone$-$\Ms$ sequence is still relatively tight (the scatter is 0.24 dex in $\sigone$, and $\sim 0.3$ dex in $\Ms$).  \cite{Fang2013} noted that the quiescent relation is offset to higher $\sigone$ at the same $\Ms$ compared to the SF relation (the offset is $\sim$0.3 dex at $\Ms=10^{10.25} \Msun$) and interpreted this as a mass-dependent density threshold for quenching.  This offset is one manifestation of the morphology-quiescence relation.  \cite{Barro2017} showed that both the SF and Q sequences evolve modestly with redshift, their zero-points decreasing by $\ltsima$ 0.3 since $z\sim 3$.

\footnotetext{Without volume correction the slope for the Q population is 0.64, with all other parameters not significantly changed.  This is more consistent with the quiescent slope reported in Fang et al. (2013), Barro et al. (2017) and Tacchella et al. (2017), none of whom perform volume weighting.  Nevertheless, the slope of the quiescent population is not used in the remainder of this study.}

Star forming galaxies are expected to increase their mass by a factor of $\sim 20$ since $z\sim 3$ (\citealp{VanDokkum2010,Behroozi2013,Moster2013,Papovich2015}) which is much larger than the scatter of the $\sigone$-$\Ms$ scaling relation and the evolution of the zero-point.  Thus \cite{Barro2017} argued that galaxies likely evolve along the SF relation which has a slope of $\sim 0.9$ at all $z$ (our best fit for the SDSS DR7 has a slope of 0.86).  Chen et al. (in preparation) demonstrate from the growth of halos that the slope of the evolutionary track of a single galaxy is expected to be closer to 0.7, \ie, shallower than the $\sigone$-$\Ms$ relation for the population as a whole.  Either way, the slope of the secular evolutionary track is predicted to be $\ltsima$ 1.  This means that the inner kpc grows more slowly than the galaxy as a whole, which is consistent with inside-out disk growth (\eg, \citealp{Fall1980,VandenBosch1998,Kepner1999,Lilly2016}).  We might therefore expect these SF galaxies to have negative age gradients (older centres), since they have built their centres in the distant past, but continue to grow their disks.

However, SF galaxies could deviate from this secular disk growing mode if they experience a compaction-like event that triggers gas inflow and the growth of the bulge/core.  A compaction event would lead to the galaxy being offset above the SF sequence of the $\sigone$-$\Ms$ relation, such that its central density is high for its $\Ms$.  
Indeed, such evolution is observed in hydrodynamical simulations, where galaxies evolve along the SF sequence in $\sigone$-$\Ms$ until a major dissipative event pushes their evolution upwards along pathways that are steeper than the SF sequence (see Fig. 19 of \citealp{Zolotov2015} and Fig. 5 of \citealp{Barro2017}).  Such compaction events are triggered by major mergers, disk instabilities or any other mechanism that brings significant amounts of gas to the centre of the galaxy resulting in enhanced star-formation in the centre (\citealp{Dekel2009,Zolotov2015,Tacchella2016}).  Although the most violent and gas rich compaction events are prevalent at high $z$, at later times gas-rich processes may also include longer-lived bar instabilities (see \citealp{Lin2017}) and mergers.  Since these lower-$z$ events are still gas-rich and dissipative, we consider them ``compaction-like'' (see \secref{discussion} for a discussion of the nature of low-$z$ ``compaction'').  Galaxies experiencing these compaction-like events have added new stars to their centres (\eg, \citealp{Ellison2018,Hall2018,Wang2018}).  Thus at the end of their steeper trajectories in the $\sigone$-$\Ms$ plane, their age gradients should be more positive than expected given secular inside-out growth.  

In light of the expected evolution of SF galaxies in the $\sigone$-$\Ms$ diagram for these two growing modes, we divided the galaxies into those with ``compact'' and ``diffuse'' cores.  The equation we use to divide the galaxies is $\log \sigone=0.86(\log \Ms-10.25)+9.1$ which is the best-fit line of the SF sequence (blue dashed line in \fig{ridgeline}), offset upward by 0.1 dex so that the Q population also has sufficient numbers of ``diffuse'' cores.  Our results are not qualitatively sensitive to reasonable variations in the placement of this line.

\begin{figure*}
\includegraphics[width=0.48\linewidth]{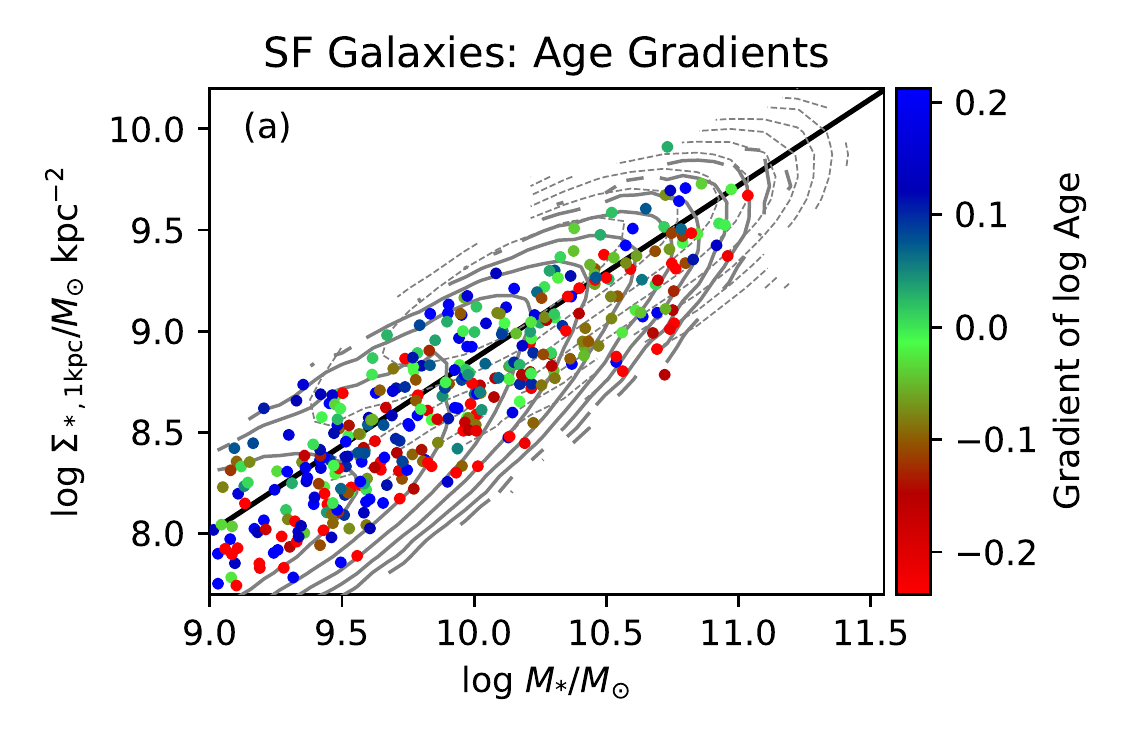}
\includegraphics[width=0.42\linewidth]{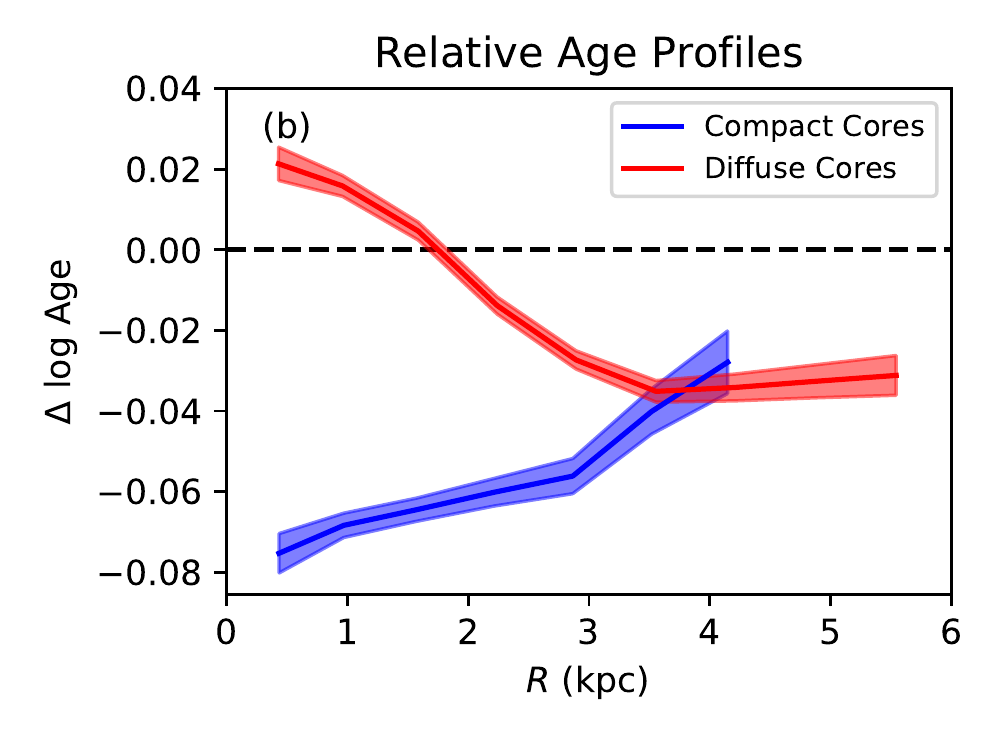}
\caption{\small (a) The age gradient as a function of $\sigone$ and $\Ms$ for SF galaxis in MaNGA (points).  The grey contours mark the SF and Q populations in the SDSS DR7 sample.  The black line divides ``compact'' from ``diffuse'' cores.  (b) The smoothed median relative age profiles for galaxies with compact (blue) and diffuse (red) cores.  The thickness of the curves is the error on the median.  Galaxies with compact cores have relatively younger centres compared to their outskirts while galaxies with lower $\sigone$ have older centres.  
} 
\label{ridgeline_age}
\end{figure*}

\begin{figure*}
\includegraphics[width=0.48\linewidth]{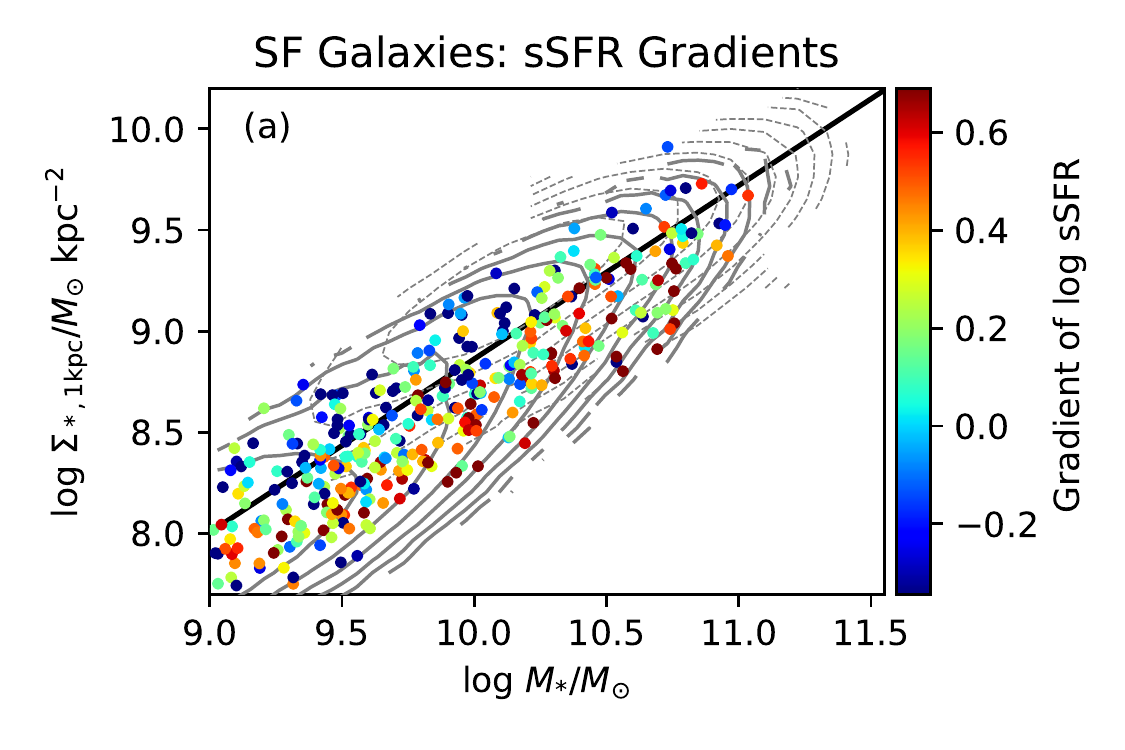}
\includegraphics[width=0.42\linewidth]{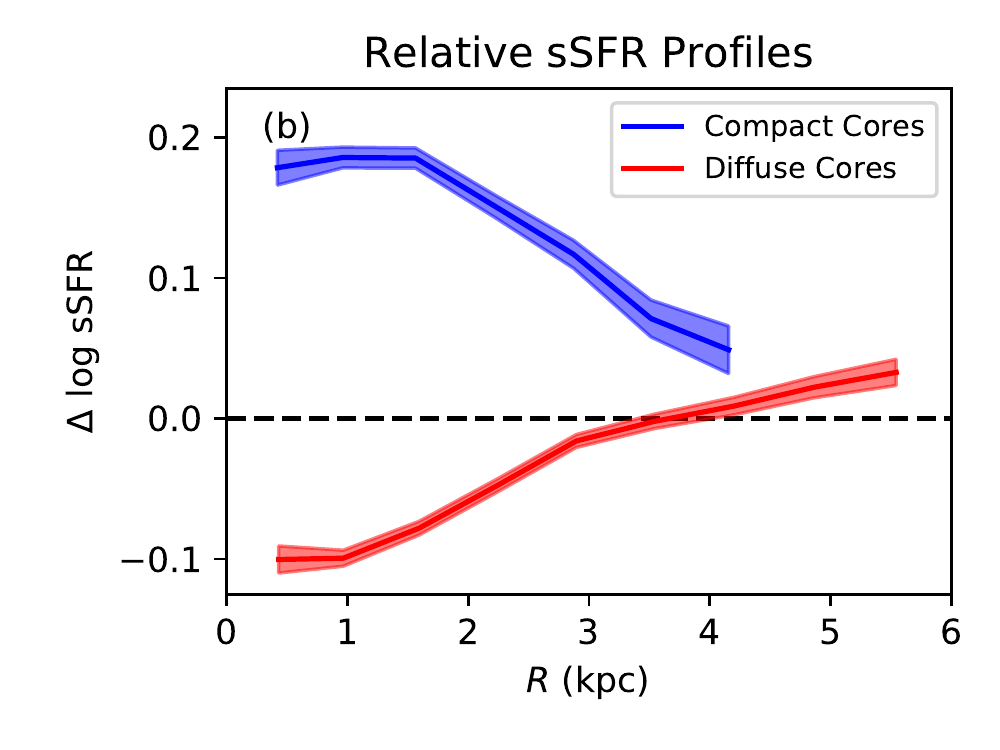}
\caption{\small  (a) The sSFR gradient as a function of $\sigone$ and $\Ms$ for SF galaxies in MaNGA (points).  The grey contours mark the SF and Q populations in the SDSS DR7 sample.  The black line divides ``compact'' from ``diffuse'' cores.  (b) The smoothed median relative sSFR profiles for galaxies with compact (blue) and diffuse (red) cores.  The thickness of the curves is the error on the median.  Galaxies with compact cores have centrally peaked sSFR while galaxies with lower $\sigone$ form stars in their outskirts. 
} 
\label{ridgeline_ssfr}
\end{figure*}

\begin{figure*}
\includegraphics[width=0.48\linewidth]{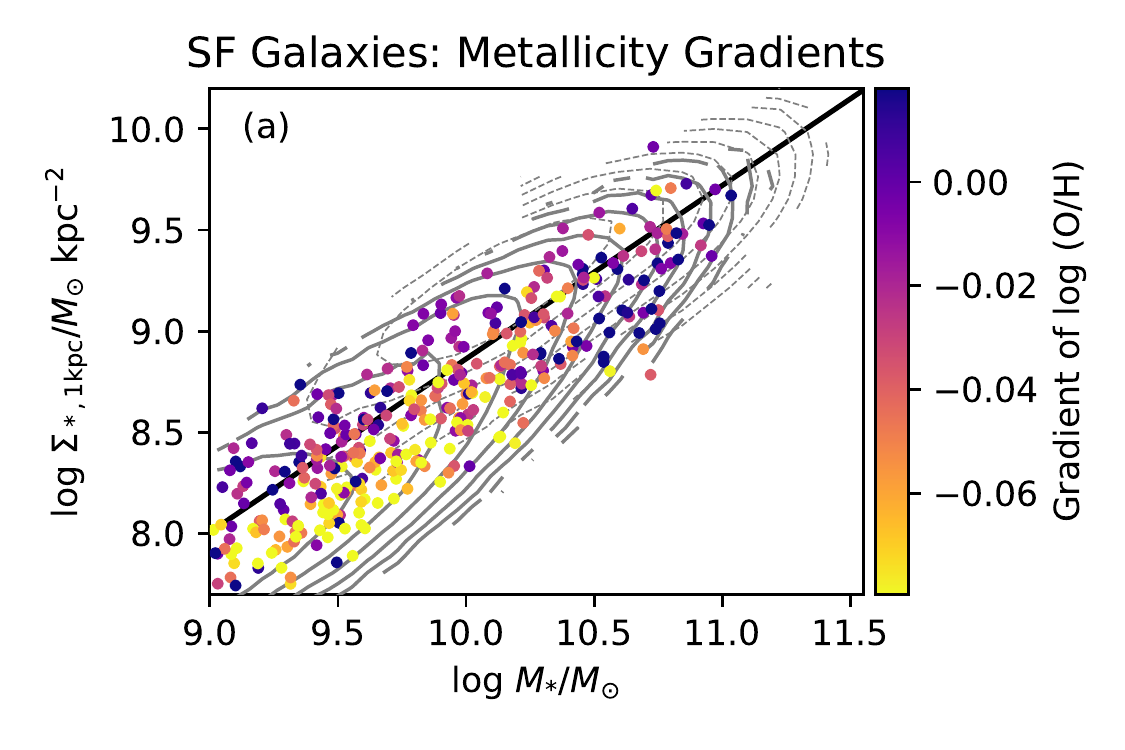}
\includegraphics[width=0.42\linewidth]{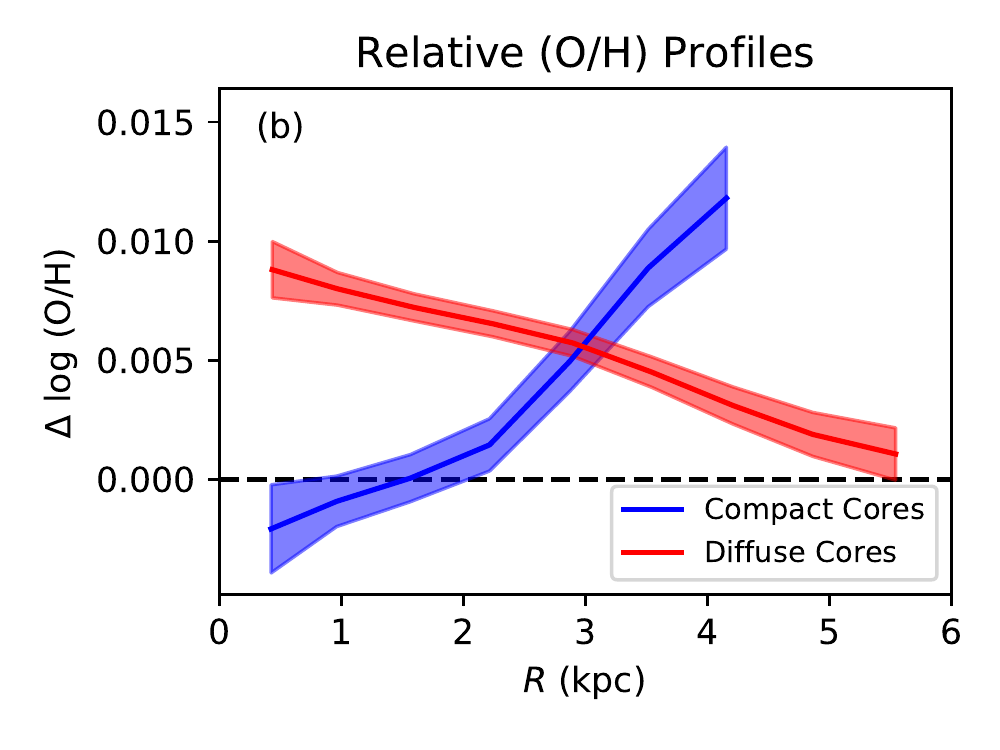}
\caption{\small (a) The gradient of 12 + log(O/H) as a function of $\sigone$ and $\Ms$ for SF galaxies in MaNGA (points).  The grey contours mark the SF and Q populations in the SDSS DR7 sample.  The black line divides ``compact'' from ``diffuse'' cores.  (b) The smoothed median relative log(O/H) profiles for galaxies with compact (blue) and diffuse (red) cores.  The thickness of the curves is the error on the median.  Galaxies with compact cores have relatively metal-poor gas in their centres while galaxies with lower $\sigone$ have O/H profiles that mildly decrease with radius. 
} 
\label{ridgeline_gasmet}
\end{figure*}

\subsection{Age Gradients and Profiles}
\label{agegradients_sf}

Here we test whether galaxies on/above the SF sequence of the $\sigone$-$\Ms$ relation have different age gradients.  To that end, we show the age gradients (\secref{graddelta}) of MaNGA galaxies as a function of the $\sigone$-$\Ms$ relation in \fig{ridgeline_age}a.   This figure shows only SF galaxies from MaNGA, with the SF and quiescent contours for the general SDSS DR7 population (reproduced from \fig{ridgeline}) in the background for reference.  The points in \fig{ridgeline_age}a show the individual MaNGA galaxies colour-coded by their age gradient.  There is significant scatter in the age gradients, but \fig{ridgeline_age}a hints
that at fixed stellar mass, star-forming galaxies 
with lower $\sigone$ tend to have negative age gradients (older centres).  The red points, for example, (those with gradients of log age $\ltsima$ -0.1) tend to lie below the black line.  This black line is the best fit to the SF sequence offset upward by 0.1 dex, \ie, our division between galaxies with ``compact'' or ``diffuse'' cores.

We caution here that the numerical scale on the colour bar in \fig{ridgeline_age}a is somewhat sensitive to the stellar population models used.  Our fiducial analysis uses the E-MILES templates.  The public age gradients from Firefly \citep{Goddard2017} are more or less consistent with our pPXF age gradients.  Using the \cite{Maraston2011} models with pPXF produces qualitatively similar trends, but the age gradients are always negative.  What is consistent between the models is that SF galaxies with denser cores have {\it more positive} age gradients than those with lower $\sigone$ at the same mass, especially at lower masses.

Given the uncertainties in the age gradients as well as the apparent mass trends, we compute enhancements and deficits in age and show their profiles in \fig{ridgeline_age}b comparing compact and diffuse cores.
These profiles quantify whether a given baxel has a younger or older age than expected given its stellar mass density, radial position, and $\Ms$ of its host galaxy (\secref{graddelta}).  A $\dlage$ of 0 means the baxel has the same log age as the average of similar baxels.
Note that although only galaxies with at least 10 good baxels (required to define a gradient) are shown on panel (a), all good baxels (criteria in \secref{agedef}) from all SF galaxies are represented in panel (b).  
The standard error on the median $\dlage$ is represented by the width of the shaded region.  Note that the offsets between the median profiles are very small, much smaller than the errors on individual age estimates.   However, we visually checked the profiles of individual galaxies (shown in Appendix B) and confirm that the median profiles accurately represent the overall radial trends of most individual galaxies. 

\fig{ridgeline_age}b shows that baxels in the centres of galaxies with compact cores are young (have negative $\dlage$) relative to the typical baxel of the same mass density, position and host galaxy mass.  Conversely, galaxies with diffuse cores are relatively older in their centres.  
The trends exhibited by the age profiles are therefore consistent with an evolutionary picture that includes both inside-out disk growth  and compaction-like processes.  
In other words, some galaxies seem to grow their disks inside-out and have relatively older centres, while others appear to experience a compaction-like event that adds new stars to their centres, increases their core density and places them on the upper part of the $\sigone$-$\Ms$ diagram.

\subsection{sSFR Gradients and Profiles}

In the previous sub-section (\fig{ridgeline_age}), we demonstrated that SF galaxies with compact cores (at fixed $\Ms$) have relatively young central ages compared to their outskirts, while SF galaxies with diffuse cores have older central ages.  In this sub-section we test whether these differences in age gradients and $\dlage$ profiles are also seen in the current star formation rates.  In \fig{ridgeline_ssfr}a, we once again show the distribution of our MaNGA sample on the $\sigone$-$\Ms$ plane, but now colour coded by their sSFR gradients. 
The corresponding profiles of relative sSFR are shown in \fig{ridgeline_ssfr}b, comparing galaxies with compact vs. diffuse cores.  Individual profiles are shown in Appendix B.  \fig{ridgeline_ssfr} is therefore analogous to \fig{ridgeline_age} except that the sSFR gradients and the profiles of $\dssfr$ are computed only from baxels that meet our star-forming criteria (\secref{sfdef} and \ref{graddelta}).  

Panel (a) of \fig{ridgeline_ssfr} shows that the sSFR gradients tend to be negative (high sSFR in the centres) for galaxies with compact cores.  These galaxies lie in the region of the $\sigone$-$\Ms$ diagram where the age gradients (\fig{ridgeline_age}a) tend to be positive, and where galaxies are expected to evolve after compaction-like events.  \Fig{ridgeline_ssfr}b confirms that the sSFR is enhanced in the centres of galaxies with dense cores.  Thus the relatively younger centres of galaxies with dense cores are experiencing centrally-concentrated star formation consistent with a compaction-like event.  Galaxies with diffuse cores have very different sSFR profiles; sSFR gradients tend to be positive (higher sSFR in the outskirts - \fig{ridgeline_ssfr}a), and their sSFR profiles are suppressed in their centres.

\subsection{O/H Gradients and Profiles}

Lastly, we show the gradients of gas phase metallicity and the corresponding profiles of relative metallicity in \fig{ridgeline_gasmet}a and b using only star-forming baxels (\secref{sfdef} and \ref{graddelta}) to study the composition of the gas feeding the star formation.  \fig{ridgeline_gasmet}a shows that the upper part of the $\sigone$-$\Ms$ SF sequence (galaxies with compact cores) is dominated by more positive gas metallicity gradients, while the lower part consists of galaxies with more negative gradients.  \fig{ridgeline_gasmet}b shows that the centres of star-forming galaxies with compact cores, where the stars are relatively young and the sSFR is enhanced, the gas fuelling the star formation is deficient in metals.  In contrast, star forming baxels in galaxies with diffuse cores have higher relative gas metallicities compared to galaxies with compact cores, and declines with radius.  Again, the offsets between the median profiles are much smaller than the errors on individual measurements.   The profiles of individual galaxies (shown in Appendix B) in reality have a wide range of offsets.  However we find that profiles of individual galaxies (shown in Appendix B) do seem to decline towards to the centres of compact galaxies and increase toward the centres for galaxies with diffuse cores.

\medskip 

Taken together, the variation in galaxies' central mass density at a fixed $\Ms$ (\fig{ridgeline}), and their contrasting profiles of age, sSFR and O/H (\mfigs{ridgeline_age}{ridgeline_gasmet}), indicate that galaxies likely do not follow a single growth pathway. At the very least, the data are consistent with two complementary routes.  On one path galaxies grow inside-out by adding stars preferentially to the outskirts.  Their total mass increases faster than the mass in their cores and they evolve along the lower part of the star-forming sequence in the $\sigone$-$\Ms$ diagram.  On the second path, some galaxies experience a compaction-like event that grows the core through centrally concentrated star formation fed by gas that is relatively low in metallicity compared to the outskirts.  These galaxies lie in the upper part of the star-forming sequence in the $\sigone$-$\Ms$ diagram since their core mass has increased.

\section{The $\sigone$-$\Ms$ Relation of the Quiescent Population}
\label{sfridgelinesec_q}

\begin{figure*}
\includegraphics[width=0.48\linewidth]{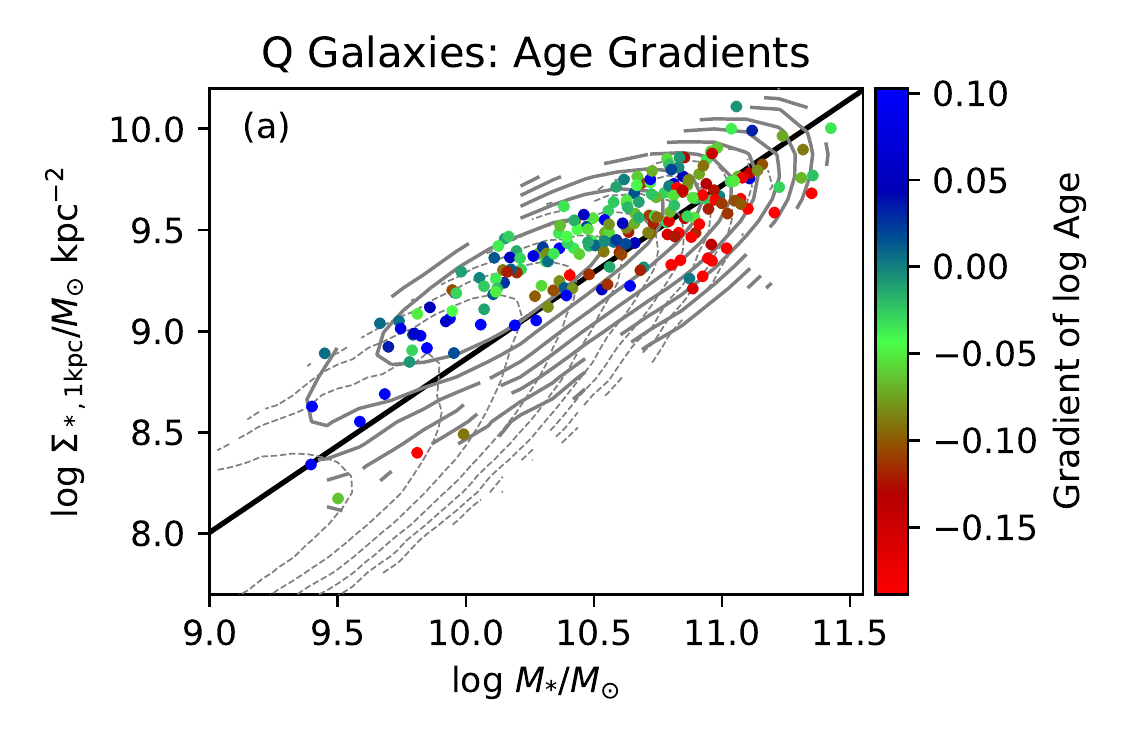}
\includegraphics[width=0.42\linewidth]{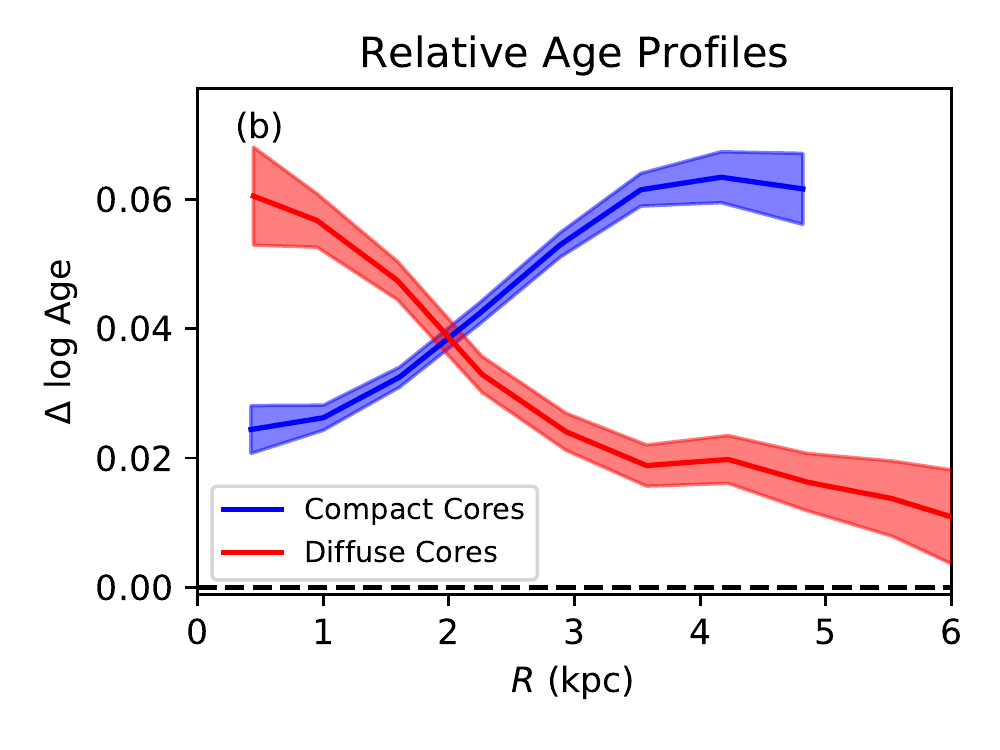}
\caption{\small (a) The age gradient as a function of $\sigone$ and $\Ms$ for the Q population in MaNGA (points).  The grey contours mark the SF and Q populations in the SDSS DR7.  The black line divides ``compact'' from ``diffuse'' cores.  (b) The smoothed median relative age profiles for galaxies with compact (blue) and diffuse (red) cores.  The thickness of the curves is the error on the median.  Quiescent galaxies with more compact cores (relative to their $\Ms$) are relatively younger in their centres than galaxies with lower $\sigone$.  
} 
\label{ridgeline_age_q}
\end{figure*}

\begin{figure*}
\includegraphics[width=0.48\linewidth]{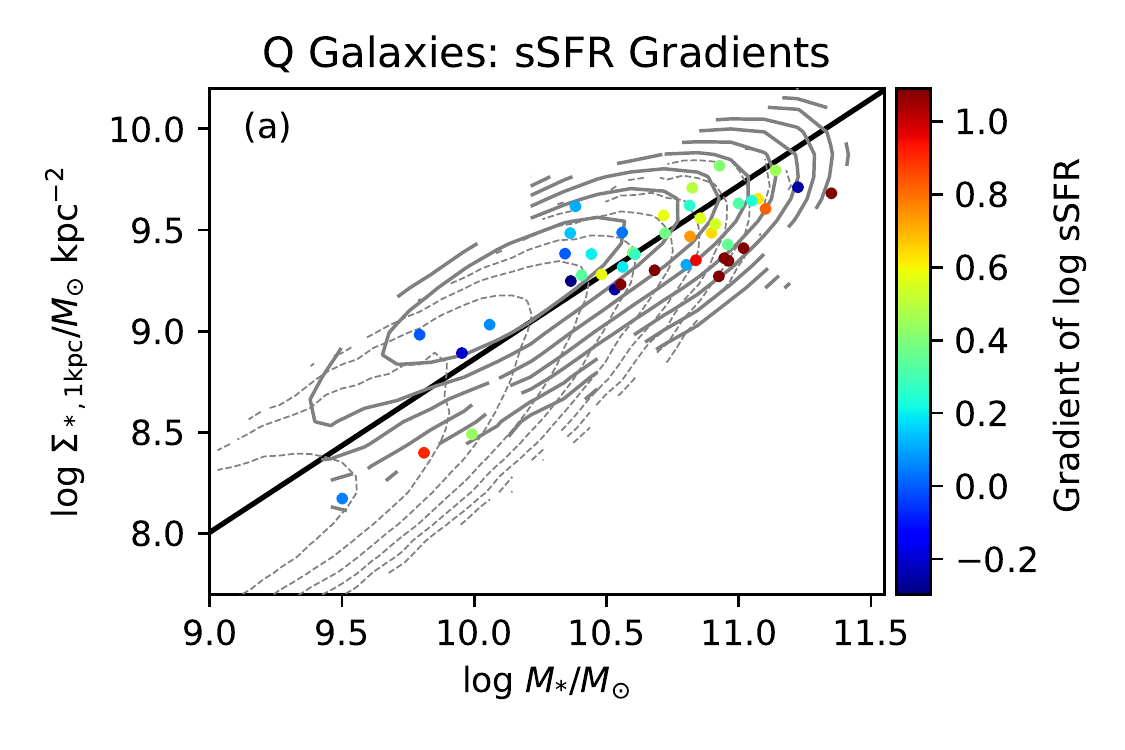}
\includegraphics[width=0.42\linewidth]{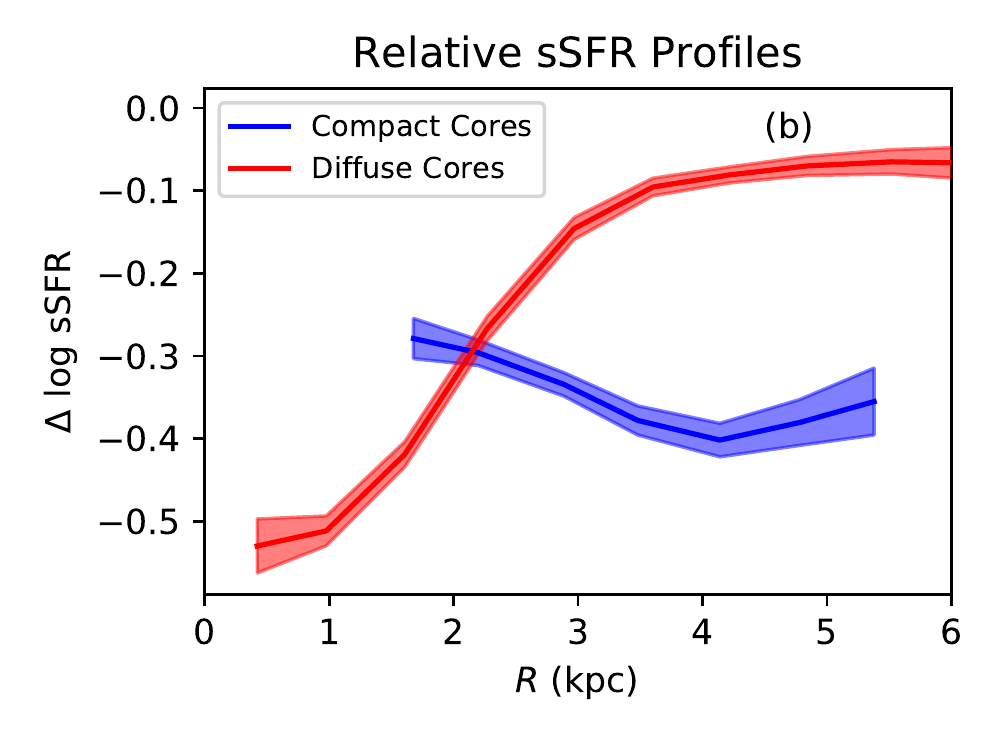}
\caption{\small (a) The sSFR gradient as a function of $\sigone$ and $\Ms$ for the Q population in MaNGA (points).  The grey contours mark the SF and Q populations in the SDSS DR7.  The black line divides ``compact'' from ``diffuse'' cores.  (b) The smoothed median relative sSFR profiles using only star-forming baxels in galaxies with compact (blue) and diffuse (red) cores.  The thickness of the curves is the error on the median.  Residual star formation in quiescent galaxies with compact cores declines outward, while in galaxies with diffuse cores, sSFR is more suppressed in the centres.} 
\label{ridgeline_ssfr_q}
\end{figure*}

\begin{figure*}
\includegraphics[width=0.48\linewidth]{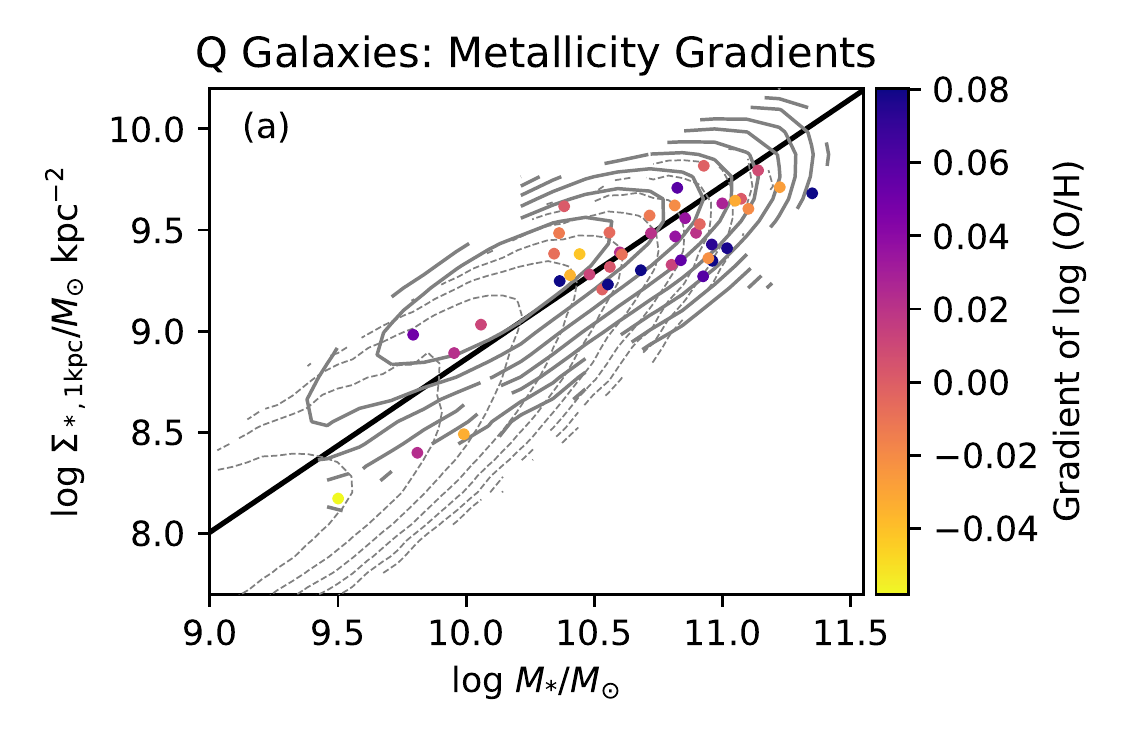}
\includegraphics[width=0.42\linewidth]{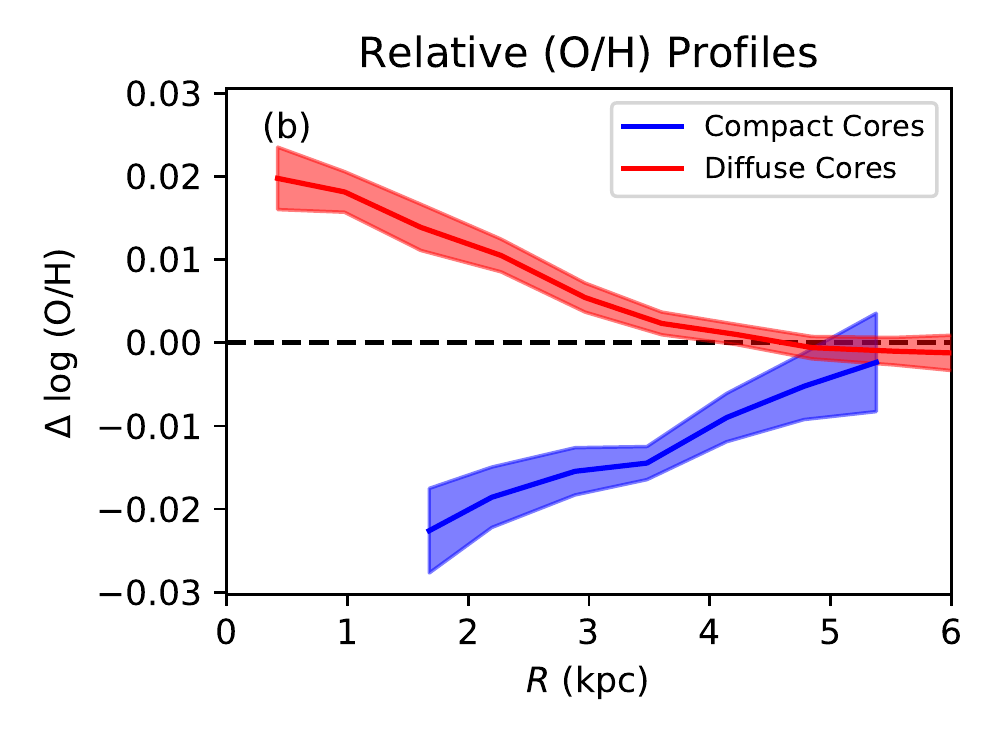}
\caption{\small (a) The gradient of 12 + log(O/H) as a function of $\sigone$ and $\Ms$ for Q galaxes in MaNGA (points).  The grey contours mark the SF and Q populations in the SDSS DR7.  The black line divides ``compact'' from ``diffuse'' cores.   (b) The smoothed median relative log(O/H) profiles for star-forming baxels in galaxies with compact (blue) and diffuse (red) cores.  The thickness of the curves is the error on the median. Compact cores are relatively metal-deficient while galaxies with diffuse cores have metal-deficient outskirts.} 
\label{ridgeline_gasmet_q}
\end{figure*}

As noted in \secref{sfridgelinesec} and shown in \fig{ridgeline}, the $\sigone$-$\Ms$ relation for quiescent galaxies lies above the relation for the star-forming galaxies.  However there is substantial overlap between the star-forming and quiescent relations.  For a galaxy to become quiescent, it must evolve along the star-forming relation until it reaches the region of $\sigone$-$\Ms$ space that is covered by the quiescent sequence, which includes a large portion of the upper part (high $\Ms$ as well as $\sigone$) of the SF relation.  Thus galaxies can potentially reach the quiescent relation via either of the growth routes discussed in \secref{sfridgelinesec}: the lower ``inside-out'' growth route, or the upper ``compaction'' route.  In this section, we investigate the signatures of both routes in the stellar ages, sSFR and metallicities of the quiescent population.  Figs. \ref{ridgeline_age_q}, \ref{ridgeline_ssfr_q} and \ref{ridgeline_gasmet_q} are analogous to Figs. \ref{ridgeline_age}, \ref{ridgeline_ssfr} and \ref{ridgeline_gasmet}, except that they plot quiescent, rather than SF galaxies.  Again, individual profiles are shown in Appendix B.  We discuss each of these figures in turn.

\fig{ridgeline_age_q} shows the gradients of stellar age for quiescent galaxies in the $\sigone$-$\Ms$ plane.  Panel (a) of this figure shows that quiescent galaxies have mostly negative age gradients (older centres).  However, we note again that the numerical scale on the colour bar is somewhat sensitive to the input stellar population models.  For example, the age gradients determined from the Firefly fitting software (using the \citealp{Maraston2011} models - see \citealp{Goddard2017}) produce strongly positive age gradients for most quiescent galaxies.  However, the {\it relative} gradients are roughly consistent among all models: galaxies with higher $\sigone$ have {\it more positive} age gradients than those with lower $\sigone$ (at fixed $\Ms$).  This is seen as a gradient in the colours of the points in \fig{ridgeline_age_q}a. 

As a complementary demonstration of the different age profiles of galaxies with different core densities, we show the mean relative age profiles in \fig{ridgeline_age_q}b comparing galaxies above (``compact'') and below (``diffuse'') the black line in panel (a), which is the same line used to divide compact and diffuse cores for the SF galaxies.  Although the line lies below the main bulk of Q galaxies, it happens to roughly separate the red points (gradient of log age $\ltsima$ -0.05) from the green and blue points (gradient of log age $\gtsima$ -0.05), suggesting a relation with the SF evolutionary paths discussed in \secref{sfridgelinesec}.  The quantity on the $y$-axis of panel (b) is again the offset in age of a baxel compared to the expected age of similar baxels.  \fig{ridgeline_age_q}a shows that quiescent galaxies with dense cores (high $\sigone$ relative to their $\Ms$) have age gradients that are more positive than those with diffuse cores.

We remind the reader that, in order to facilitate comparison with other studies, quiescent galaxies were defined using the criterion log sSFR/yr$^{-1} < -11$, where the sSFR is the global value (not the baxel value) from the widely used MPA-JHU DR7 catalogue (\citealp{Brinchmann2004}).  Despite their overall ``quiescent'' nature, over half of these galaxies have residual emission lines from individual star-forming regions although their total sSFR inferred from these lines is consistently low.  Galaxies that fulfill our 10-baxel requirement for a gradient in sSFR and (O/H) (\secref{graddelta}) are about 
$\sim 20\%$ of quiescent galaxies.  The gradients and profiles of sSFR and gas metallicity are shown in \twofigs{ridgeline_ssfr_q}{ridgeline_gasmet_q}.

For consistency with previous figures we show both the gradients of individual galaxies as well as the median profiles, although the gradient plots have too few points to draw meaningful conclusions.  However, the profile plots do provide some interesting information.  \fig{ridgeline_ssfr_q}b shows that for quiescent galaxies with residual star formation, galaxies with compact cores have suppressed sSFR everywhere, as their $\dssfr$ values are all below 0.  Their profiles have negative slope which means sSFR is suppressed more in the outskirts.  In fact their slope is not very different from the slope of the $\dssfr$ profiles of the SF galaxies (blue curve in \fig{ridgeline_ssfr}b) even if the overall normalization is lower.  Note that the centres of galaxies with compact cores are generally missing star-forming baxels.  This is not necessarily because these baxels do not have star formation.  Their emission lines seem to be dominated by AGN-like emission.  However the individual profiles (shown in Appendix B) do seem to turn over towards the centres indicating that sSFR in the centres may indeed be low.  
In contrast, quiescent galaxies with diffuse cores have suppressed sSFR in their centres compared to their outskirts.  Comparing the red curves in Figs.  \ref{ridgeline_ssfr}b and \ref{ridgeline_ssfr_q}b, the decline in the mean sSFR profile towards the centre is steeper for Q galaxies than for SF galaxies if they have diffuse cores.

\fig{ridgeline_gasmet_q}b shows that the gas metallicities of the same compact quiescent galaxies are relatively deficient while the centres of galaxies with diffuse cores are enhanced in metallicity.  However from looking at the individual profiles (Appendix B), we caution that the median profiles may be skewed by one or two outliers.  While we show them for completeness, we do not make any broad conclusions from these profiles.

Our findings indicate that quiescent galaxies with different core densities have different gradients and profiles of stellar age, sSFR and possibly also gas metallicity.  
The presence of both positive and negative profiles in age, sSFR and O/H amongst the quiescent population may be pointing to different evolutionary pathways that resulted in their quiescence.  We now turn to a discussion of these possibilities.

\section{Discussion: the Growth and Quenching of Galaxies}
\label{discussion}

Our study was motivated by the question of distinguishing between core-building ``compaction''-like events (\citealp{Zolotov2015}) and progenitor effects of secular inside-out growth (\citealp{Lilly2016}) as the reason for the morphology-quiescence relation.  Through our study of the age, sSFR and gas phase metallicity gradients and radial profiles as a function of position on the $\sigone$-$\Ms$ diagram, we have uncovered evidence pointing to at least two pathways of growth consistent with 1) compaction-like core-building events and 2) secular inside-out growth by global star formation.  The cores of compact (high $\sigone$ at given $\Ms$) star-forming galaxies are relatively young, enhanced in sSFR, and deficient in metals. In contrast, the centres of galaxies with diffuse cores (low $\sigone$ for their $\Ms$) are characterized by relatively older stellar ages, lower sSFR and higher O/H.
We showed that the quiescent galaxies also have age, sSFR and gas metallicity profiles that depend on their central stellar mass densities.  Like the SF population, quiescent galaxies with compact cores differ from those with diffuse cores in their stellar ages, sSFR and probably also gas metallicities (when gas is present).  Our results suggest that both modes of growth (secular and compaction) contribute to the quiescent population and that the morphology-quiescence relation results from both modes.  

Multiple {\it quenching} modes for isolated or central galaxies have been suggested in the literature before (\citealp{Barro2013,Yesuf2014,Schawinski2014,Woo2015,Maltby2018,Wu2018}).  These quenching modes were proposed as fast and slow modes (or early/late), depending on galaxy morphology/structure.  These quenching modes could be related to the compaction-like and secular growing modes argued for here.  Evidence for multiple {\it growing} modes are only now emerging in recent research.  \cite{Wang2017} selected a sample of 77 galaxies in MaNGA with positive D4000 gradients (\ie, with young centres) and elevated central EW($\Ha$) (elevated central SFR) and found that these are smaller, more concentrated and more bulge-dominated than a control sample of star-forming galaxies.  \cite{Wang2017} argued that these galaxies are undergoing rapid bulge growth in contrast to other star-forming galaxies. 
\cite{Liu2018} divided the galaxies in MaNGA essentially by age gradient (more precisely, D4000 gradient) and found that galaxies populate different spatially resolved SFR-$\sigs$ relations.  Galaxies with older centres have shallower SFR-$\sigs$ relations than those with younger centres (see also \citealp{Hall2018}).  In other words, galaxies with older centres have suppressed SFR in spaxels of the highest $\sigs$ (\ie, in their centres) compared with galaxies with younger centres.  Furthermore, \cite{Wu2018} found that despite a general correlation between age (via D4000 and EW(H$\delta$)) and galaxy size, young post-starburst galaxies are much smaller than this trend would suggest, suggesting multiple evolutionary pathways.  Our findings are consistent with these studies and have explicitly linked these trends of age and SFR gradients with the compactness of the cores, which presumably grow via these growing modes.

If our interpretation of at least two growing modes is correct, then we can make a few inferences about quenching which we unpack below.  Since our division between galaxies with compact and diffuse cores was the same in our analyses of both SF and Q galaxies, for the purposes of the ensuing discussion, we will assume that the position on the $\sigone$-$\Ms$ diagram for the Q galaxies (\ie, having compact vs. diffuse cores) is the end point of the two growing pathways.  In essence, this means that neither $\sigone$ nor $\Ms$ evolves much after quenching.  This is not an unreasonable assumption given that the quiescent sequence significantly overlaps the SF sequence in this diagram.  The density within a scale radius ($\Sigma_{\rm eff}$) can easily decrease through minor mergers which add stars to galaxy outskirts (\eg, \citealp{Naab2009,Hopkins2010a,Bedorf2013}).  However $\sigone$ does not easily decrease (\citealp{Hopkins2010a,Zolotov2015,Barro2017}), especially in isolated environments.  
Furthermore, any movement towards the ``diffuse core'' part of the diagram after compaction would likely require an increase in $\Ms$ by at least a factor of two (a horizontal  movement), which may be difficult to achieve through merging without also increasing $\sigone$.  

\subsection{The Compaction-like Evolutionary Track}

In the growing mode that involves a compaction-like event, low-metallicity gas is funnelled to the central regions of a galaxy.  This gas fuels a centrally concentrated burst of star formation (\fig{ridgeline_ssfr}b) that increases their $\sigone$ faster than their $\Ms$, bringing them to the upper part of the $\sigone$-$\Ms$ sequence (\fig{ridgeline}).  
Classical compaction events resulting from violent disk instabilities are a high-$z$ phenomenon \citep{Zolotov2015}, triggered by low-metallicity cosmic streams \citep{Dekel2009,Cresci2010}.  However, our study is limited to $z\sim 0$ galaxies (so we have been calling this track ``compaction-like'').   Furthermore, we have observed centrally concentrated star formation in the upper $\sigone$ track occurring in the last few Myr (through $\Ha$ emission).  What is the nature of this low-$z$ phenomenon characterized by the gradients and profiles of stellar age, sSFR and gas metallicity?  

One possibility is galaxy mergers.  The relatively low metallicity in the gas of these galaxies (\fig{ridgeline_gasmet}b) is reminiscent of what has been seen in galaxies expected to be currently under-going a merging event (low central O/H, and/or flatter gradients: \citealp{Kewley2010,Scudder2012,Ellison2013,Thorp2019}), as well as galaxies with enhanced central star formation rate (\citealp{Ellison2018}).  Furthermore, $\sigone$ correlates tightly with the central velocity dispersion at fixed $\Ms$ (\citealp{Fang2013}) which may suggest a more active merger history in the upper part of the $\sigone$-$\Ms$ sequence compared to the lower part.  However, major mergers are rare in the local universe \citep{Darg2010}, and their importance in galaxy evolution remains disputed (e.g. \citealp{Kaviraj2015}).  Nevertheless, their cumulative impact can be significant.

The recent studies of \cite{Lin2017} and \cite{Chown2019} hint at another possible low-$z$ ``compaction'' mechanism.  These authors study the inner regions of $\sim$60 star-forming disks galaxies in CALIFA and find that the centres of 17-19 have a drop in D4000 (younger ages) and an enhancement of EW($\Ha$) (sSFR), similar to our findings for galaxies on the upper part of the $\sigone$-$\Ms$ diagram.  They found that most of these are barred galaxies, suggesting that low-$z$ ``compaction'' may involve bar instabilities.   However, simulations predict {\it higher} gas phase metallicities in barred galaxies, especially if they are accompanied by central star formation \citep{Friedli1994,Friedli1995,Martel2013,Martel2018}.  Indeed \cite{Ellison2011} found that although barred galaxies in the SDSS have elevated SFR in their centres, their gas metallicities are higher than unbarred galaxies in contrast to our findings of lower metallicities in the centres of galaxies on the compaction track.  \cite{Ellison2011} speculated that their findings implied that bars are a long-lived phenomenon.  Therefore they may be a completely different population from our MaNGA galaxies with high $\sigone$.
Clearly, further study is required to ascertain the nature of low-$z$ compaction.

\begin{figure}
\includegraphics[width=\linewidth]{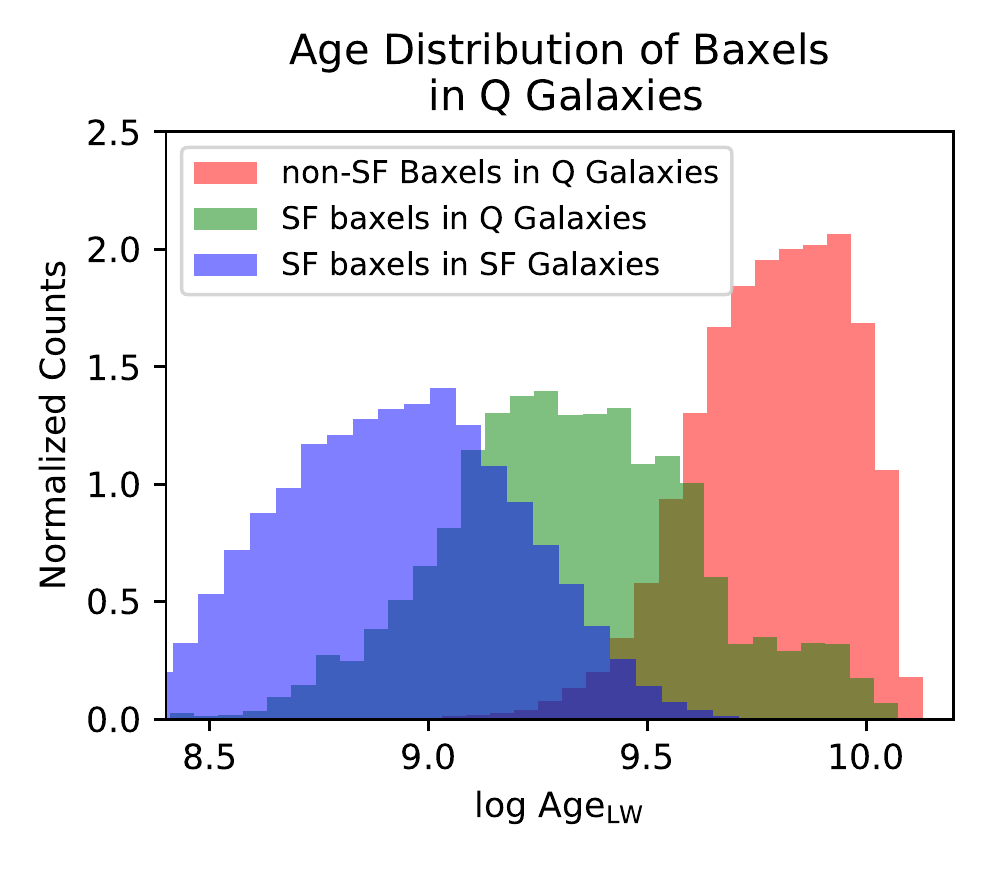}
\caption{\small The distribution of light-weighted mean stellar age for baxels inside Q galaxies.  The red histogram shows all baxels in the Q population, while the blue histogram shows only the star-forming baxels (criteria in \secref{sfdef} in the Q population.  Baxels in the Q population that fulfill the star-forming criteria are on average younger than the other baxels. } 
\label{age_q_lw}
\end{figure}

How is the compaction-like growing mode related to quenching?  It is perhaps a large leap to assume that the galaxies on the upper track of the SF sequence end up on the upper track of the Q sequence.  After all, most galaxies on the Q sequence are much older ($\sim 8-10$ Gyr) than galaxies in the SF sequence ($\sim 2-6$ Gyr), and their evolutionary pathways will not in general be the same.  In order to connect the growing modes with quenching, we require a population of galaxies that are quenching now.  The ``quiescent'' galaxies that contain baxels with residual emission lines are good candidates for such quenching galaxies (\ie,  the ones shown in \fig{ridgeline_ssfr_q}b and \fig{ridgeline_gasmet_q}b).  The typical light-weighted age of the star-forming baxels in quiescent galaxies (1.9 Gyr) is only about a Gyr older than the typical star-forming baxel (800 Myr) in SF galaxies, suggesting recent quenching.   This is in contrast to the typical light-weighted age of $\sim 6$ Gyr for non-star-forming baxels in quiescent galaxies.  We checked that the Q galaxies with and without star-forming baxels have similar mass distributions.  \Fig{age_q_lw} compares the distribution of light-weighted ages for star-forming baxels in SF and Q galaxies.
Quiescent galaxies with star-forming baxels are also likely not rejuvenating since hydrodynamical simulations predict that rejuvenation is rare, affecting only a few percent of quiescent galaxies (\citealp{Nelson2018}). 
If we assume that Q galaxies with star-forming baxels are in fact quenching, and not rejuvenating, then their relative sSFR profiles (\fig{ridgeline_ssfr_q}b) provide some clues as to how they quench.  For these quiescent galaxies, the mean $\dssfr$ profile is naturally much lower than that of the SF galaxies with compact cores (compare the blue curves of Figs \ref{ridgeline_ssfr}b and \ref{ridgeline_ssfr_q}b).  Both the SF and Q compact galaxies have decreasing $\dssfr$ profiles with radius, but the Q compact galaxies have missing baxels in their centres.  The individual profiles of Q compact galaxies seem even to turn over in sSFR towards their centres.  
If these profiles are linked as an evolutionary sequence, then 
the suppression of star formation after the compaction-like track seems to simultaneously suppress the centres quickly, but also suppresses the whole galaxy such that the declining shape of the sSFR profiles, is roughly preserved.  
Our interpretation is also consistent with the findings of \cite{Wang2017} pointing to the rapid bulge growth of galaxies with elevated central SFR and younger centres.

One popular possibility for quenching galaxies on the compaction track is AGN feedback (\citealp{Bell2008,Bluck2014,Terrazas2016,Terrazas2017,Kocevski2017}).  AGN require a fuel source, and this is naturally provided by the same dissipative compaction-like processes which funnel gas toward the centres of galaxies (gas-rich mergers and gravitational disc instabilities), increasing their $\sigone$.
The higher velocity dispersions in galaxies with higher $\sigone$ (\citealp{Fang2013}) may point to the importance of black hole growth (via the $M_{\rm BH}-\sigma$ relation - \citealp{Kormendy2013}) during the compaction event, and then later for quenching (as argued for example by \citealp{Bluck2016}).   
AGN would make sense of the missing central sSFR in \fig{ridgeline_ssfr_q}b (blue curve).  

\begin{figure}
\includegraphics[width=\linewidth]{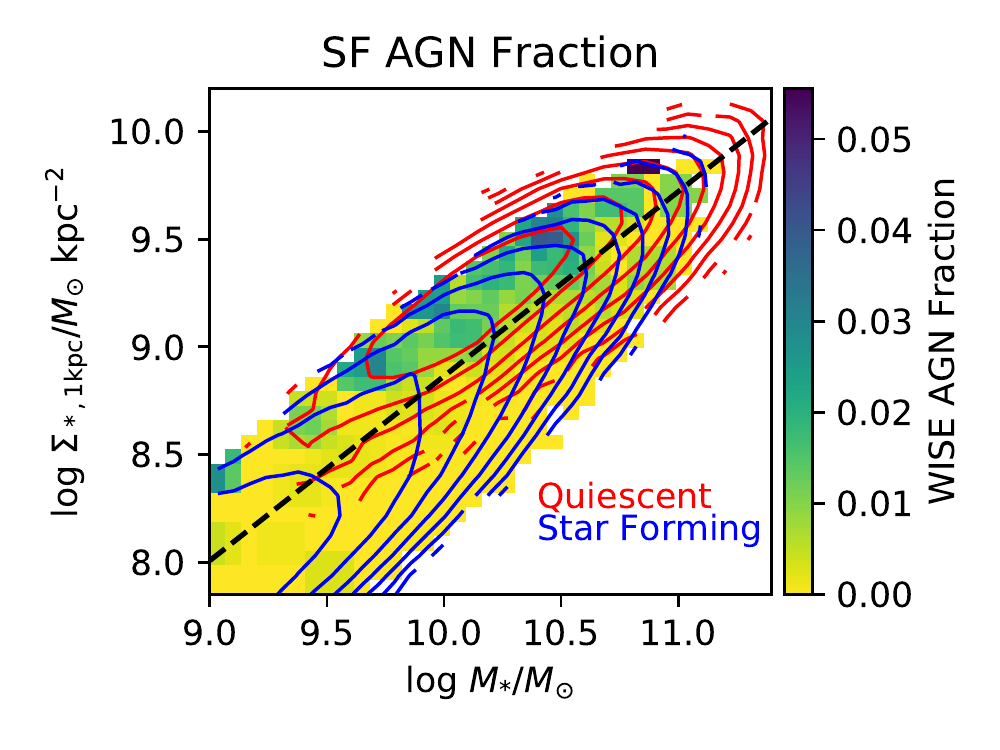}
\caption{\small The $\sigone$-$\Ms$ diagram colour-coded with the WISE-detected AGN fraction for SF  galaxies. WISE detections are matched to SDSS DR7 objects if the angular separation is less than 6''.  The AGN fraction is then computed from the fraction of WISE detections with $W1$-$W2$ colour greater than 0.77, which is the 75\% completeness criterion of \citet{Assef2018}.  The black dashed line indicates our division between ``compact'' and ``diffuse'' cores.  The WISE-detected AGN fraction peaks in the upper region of the $\sigone$-$\Ms$ diagram (galaxies with compact cores), where the profiles of stellar age, sSFR and gas metallicity are characteristic of compaction-like events.} 
\label{agnfrac}
\end{figure}

In fact, \cite{Kocevski2017} showed that for galaxies at $z\sim 2$, the X-ray detected AGN fraction peaks for galaxies of high sSFR and high $\sigone$, exactly in the region of the sSFR-$\sigone$ diagram where energy injection is needed to quench galaxies (their Fig. 8 uses U-V colour instead of sSFR).  This motivated the question of whether the AGN fraction peaks in the ($z\sim 0$) $\sigone$-$\Ms$ diagram where we propose galaxies evolve due to compaction-like events.  We show this fraction as a function of $\sigone$-$\Ms$ plane in \fig{agnfrac} using WISE-detected AGN for SF galaxies in the SDSS DR7 sample.  WISE detections are matched to SDSS DR7 objects if the angular separation is less than 6''.  After matching, we then applied the same redshift, seeing, environment, inclination and mass cuts described in \secref{sample}, yielding a sample of $\sim 29000$ galaxies.  We define the AGN fraction to be the fraction of WISE detections with $W1$-$W2$ colour greater than 0.77, which is the 75\% completeness criterion of \cite{Assef2018}.  We show WISE-selected AGN rather than emission-line AGN because the former method detects AGN obscured by dust, which is more prevalent in processes such as mergers (\citealp{Mateos2013,Blecha2018}).

\fig{agnfrac} shows that the WISE-selected AGN fraction for SF galaxies peaks in the upper regions of the $\sigone$-$\Ms$ plot, \ie, with high $\sigone$.  This is also where the centres of galaxies are relatively young, sSFR is enhanced, and the gas metallicity is low, pointing to a possible connection between AGN and the compaction-like evolutionary path.  
WISE AGN tend to correspond to galaxies with high SFR (\citealp{Ellison2016,Cowley2016,Azadi2017}), indicating that AGN and high SFR might be tracing the same process.   WISE-selected AGN have been associated with mergers (\citealp{Satyapal2014,Weston2017}; Ellison et al 2019, submitted).
A centrally concentrated boost of SFR tracing elevated nuclear accretion as may happen in mergers would be exemplary of compaction-like core-building.  The energetic phase of AGN, \ie, the phase most likely to be detected by WISE (\citealp{Mateos2013}) is short-lived (\eg, \citealp{Haehnelt1993}), but 
their expanding radio bubbles may continue to keep gas hot and prevent further star formation (\citealp{Croton2006}). 
Although correlation does not always imply causation, the considerations discussed here make AGN a natural candidate for the mechanism that suppresses sSFR in these galaxies.

\subsection{The Secular Disk-Growing Mode}

When disks are growing inside-out through the successive addition of gas in ever-growing exponential disks (or growing angular momentum), sSFR is highest in the outskirts (\fig{ridgeline_ssfr}).  $\Ms$ grows faster than $\sigone$ and galaxies evolve along the lower part of the $\sigone$-$\Ms$ relation (\fig{ridgeline}).  Gas-phase metallicity is higher in the centres than in the outskirts (\fig{ridgeline_gasmet}a) indicating that mixing is relatively unimportant.  When these galaxies reach the quenched region of the $\sigone$-$\Ms$ diagram, sSFR is suppressed most strongly in the centres.  This is suggested by the steeper sSFR profiles of the quiescent diffuse galaxies (compare the red curves in Figs \ref{ridgeline_ssfr}b and \ref{ridgeline_ssfr_q}b), assuming that quiescent galaxies with star-forming baxels are quenching today.  In other words, the quenching process after secular disk growth seems to occur ``inside-out''.  This is consistent with several other observations of inside-out quenching (\citealp{Tacchella2015a,GonzalezDelgado2016,Ellison2018,Tacchella2018,Belfiore2018,Liu2018}).  
The quenching seems to have occurred more gradually so as to allow more residual star-forming baxels to still remain in the centres, compared to the galaxies with compact cores.  
Though we have not explicitly studied timescales here, the secular disk growing mode is expected to be a quieter and longer-lived mode than the compaction-like track.  \cite{Wang2018} find that the depletion time is twice as long for galaxies with low $\sigone$ than galaxies with higher $\sigone$ (at fixed $\Ms$) supporting the slower evolution of the inside-out mode.  

Just as in the compaction case, the quenching mechanism after the inside-out growing mode could very well be AGN.  The central black hole can still accrete (slowly) as the disk grows, until a critical mass is reached (\eg, \citealp{Terrazas2016}, Chen et al., in prep.) triggering perhaps the ``radio/maintenance mode'' (\citealp{Croton2006}) of AGN in a hot halo.  Regardless of the mechanism, the gas in the centres is likely depleted/removed first.  In order to constrain scenarios further, measurements are needed of radial profiles of the depletion time, and/or the fraction of low-luminosity AGN, as a function of position on the $\sigone$-$\Ms$ diagram.

If the quiescent galaxies with lower $\sigone$ are indeed the end point of the evolution of the inside-out disk growth track, the fact that these quiescent galaxies are younger than those with higher $\sigone$ indicates that this quenching track may be more dominant today than the compaction track.  Yet, the number of quiescent galaxies with at least 10 star-forming baxels is roughly the same in the low- and high-$\sigone$ regions of the quiescent $\sigone$-$\Ms$ diagram (\fig{ridgeline_ssfr_q}a).   These findings imply that though the compaction track and the secular track may have respectively decreased and increased in importance over time, the quenching that occurs after the compaction track still continues to quench about half of galaxies today.

\subsection{Caveats}
\label{caveats}

We have interpreted our results as indicating (at least) two galaxy growing modes, both of which contribute to the morphology-quiescence relation.  However, one of the difficulties in inferring evolution from today's populations is that we observe a snapshot that includes the end result of recent evolution as well as quenching that occurred long ago.  The quiescent population today is mostly very old ($\sim$ 8-10 Gyr) and their evolutionary pathways will not in general be the same as those by which galaxies evolve today.

For example, consider the following alternative interpretation of our results: galaxies with high $\sigone$ attained their core densities long ago by some mechanism unrelated to today's SFR (e.g., high-$z$ compaction).  Galaxies with denser cores tend to have more gas flows towards their centres, producing the observed centrally concentrated sSFR profiles, and also the observed higher AGN prevalence.  Thus it is the core density that determines how a galaxy grows, rather than the growing mode that determines the core density.  In order to test this hypothesis, ideally we would measure the fraction of mass in the core that was formed recently, for example, in the last Gyr.  Even better would be the full star-formation history.  In principle, one could derive star-formation histories using the weights from pPXF, but unfortunately our tests have shown that these histories are unreliable.  
Therefore we cannot rule out this hypothesis.  However it is not clear what physical mechanism would cause gas inflows to occur preferentially in galaxies with high $\sigone$ at a given $\Ms$ to produce a relative enhancement in $\dssfr$, where the dependence of $\Sigma_{\rm SFR}$ on $\sigs$ has been removed.  
The other direction of causality, \ie, that gas inflows cause an increase in $\sigone$, is more straightforward.

Our results and interpretation rely in part on the measurement of stellar ages, which is notoriously difficult and prone to degeneracies and large errors (\eg, \citealp{Conroy2013}).  Indeed, we have found that age gradients, including their {\it sign}, are disturbingly inconsistent between fitting codes and different SSP templates.  For example, the mass-weighted stellar age gradients computed by \cite{Goddard2017} using the Firefly fitting code and the \cite{Maraston2011} templates are strongly positive for all quiescent galaxies regardless of $\sigone$.  \cite{Goddard2017} also cite other long-slit studies which find positive age gradients for early type galaxies.  This is in contrast to the mostly negative gradients found here, and also seen in early-type galaxies, for example, by \cite{Petty2013} (using the FSPS models of \citealp{Conroy2010}), by \cite{Li2018} (using pPXF and the MILES models) and by \cite{Garcia-Benito2017,Wang2017} (using STARLIGHT).   We have found that stellar ages tend to more or less agree in the bright inner galactic regions, and that the inconsistencies seem to be in the outskirts where the S/N is lowest.  

Discussing the merits and weaknesses of fitting codes and template models is beyond the scope of this work.  However, when we compared the public age gradients from Firefly (using the \citealp{Maraston2011} templates) with those from pPXF, what we have found to be robust is that galaxies with compact cores have different age profiles than those with diffuse cores, and that the difference is that the former have {\it more positive} gradients than the latter. 
If secular inside-out growth is the default growing mode of galaxies, then most age gradients should be negative (older stars in the centre).  Compaction-like events would then add new stars preferentially in the centres.  Regardless if such events render the overall age gradients to be positive or still slightly negative, the profiles of {\it relative} age should capture the effect of these dissipative events.

Furthermore, we have also presented gradients in sSFR and gas metallicities which are derived from emission line measurements.  These are robust even if the underlying stellar populations that are subtracted vary between stellar population models.  The disadvantage to emission line measurements is that they are not present in all areas of every galaxy since many galaxies are devoid of gas.  However we have found that when present with sufficient S/N, the emission lines tell a consistent story with the relative ages.  

We note that much of our discussion and interpretation has assumed that once stars are formed in situ, their positions are not significantly altered such that profiles of stellar age accurately reflect where the bulk of old and young stars were formed.  The obvious events that might disturb the distribution of stars are radial stellar migration and major mergers.  However, we expect that the primary effect of major mergers is to mix the orbits of pre-merger stars so as to wipe out pre-existing gradients (\citealp{White1980,DiMatteo2009}).  The fact that we see intact gradients which correlate with $\sigone$ and $\Ms$ in the way expected of dissipative compaction-like events and slower inside-out disk growth may actually attest to the lesser significance of mergers and/or migration in the life of galaxies than the events which led to their gradients.  
However further study is necessary to establish the importance of migration or mergers on the distribution of stellar ages, sSFR and gas metallicities in star forming and quiescent galaxies.

\section{Summary}
\label{summary}

We have measured the stellar ages, sSFR and gas-phase metallicities for galaxies from the SDSS DR14 MaNGA survey.  By utilizing SDSS DR7 imaging to compute the total and central stellar masses for these galaxies, we studied the behaviour of  the gradients and average profiles of stellar age, sSFR and O/H as a function of total mass $\Ms$ and the stellar surface density within 1 kpc $\sigone$.  Our aim is to determine whether core-building ``compaction''-like processes or progenitor effects resulting from secular disk growth are responsible for the link between quiescence and the presence of a dense stellar bulge (the morphology-quiescence relation).  Our conclusions can be summarized as follows:

\begin{enumerate}
  \item The gradients of stellar age, sSFR and O/H for star forming galaxies depend on position in the $\sigone$-$\Ms$ diagram.  Galaxies on the lower part of the $\sigone$-$\Ms$ relation (those with ``diffuse'' cores) have centres that are old, depressed in sSFR and enriched in metals.  Galaxies on the upper part of the $\sigone$-$\Ms$ relation (those with ``compact'' cores) have  centres that are young, elevated in sSFR and metal-deficient compared to their outskirts (\mfigs{ridgeline_age}{ridgeline_gasmet}).  {\it This is consistent with an evolutionary picture that includes both ``inside-out'' secular disk growth and dissipative ``compaction''-like core-building processes.}  The former grows galaxies along the $\sigone$-$\Ms$ relation through galaxy-wide star formation. The latter brings galaxies to the upper $\sigone$-$\Ms$ relation via a growth path that is steeper than the slope of the main $\sigone$-$\Ms$ relation for SF galaxies.  
 \item For quiescent galaxies, the profiles of age, sSFR and O/H (when measurable) also differ between those with compact cores and those with diffuse cores (\mfigs{ridgeline_age_q}{ridgeline_gasmet_q}).   Some residual star-forming gas is present in about 
20\% of quiescent galaxies.   If we assume that these galaxies are currently quenching or are recently quenched, then at least today, quenching retains the signatures of the two growing modes.  In other words, both the inside-out growth and compaction-like growing modes contribute to the quiescent population, and the morphology-quiescence relation results from at least both these modes. 
 \item If the quiescent galaxies represent the end point of the two evolutionary tracks, then the sSFR profiles imply that: a) galaxies that quench after the compaction-like track quench uniformly across most of the galaxy, with the exception perhaps of a more immediate quenching in the centres; and b) galaxies that quench after secular disk growth seem to suppress the SFR with a strong radial dependence, suggesting a more gradual outward moving quenching.
 \item The WISE-selected AGN fraction peaks in the upper region of the $\sigone$-$\Ms$ relation where galaxies are expected to arrive after compaction-like evolution (\fig{agnfrac}).  This is consistent with compaction-like events leading to the triggering of AGN, which may be responsible for the quenching of galaxies on this track.
 \item About half of quiescent galaxies with residual star-forming gas are in the lower part of the $\sigone$-$\Ms$ relation for quiescent galaxies, \ie, those that have the signatures of the secular inside-out disk growth track.  The compaction track may be more dominant at higher $z$ while the secular mode may be increasing in importance.  However, if the presence of gas indicates recent quenching, then the compaction track still continues to quench about half of recently quenched galaxies today.   
\end{enumerate}

From these results we conclude that star forming galaxies grow in $\sigone$ and $\Ms$ via at least two modes: a dissipative compaction-like core-building mechanism as well as inside-out galaxy-wide star formation, and that these two scenarios need not be mutually exclusive.  Both modes seem to contribute to the morphology-quiescence relation, with varying importance over time.  Compaction-like events are fed with low-metallicity gas, and probably happen on top of secular disk growth, causing them to deviate from the lower track of the $\sigone$-$\Ms$ diagram.

\section*{Acknowledgements}
We acknowledge the helpful and stimulating discussions with 
Trystyn Berg,
Asa Bluck, 
Connor Bottrell,
Michele Cappellari,
Avishai Dekel,
Sandra Faber, 
Maan Hani,
David Koo, 
Trevor Mendel,
Jorge Moreno,
Masato Onodera,
David Patton,
Luc Simard,
and 
Mallory Thorp.
SLE gratefully acknowledges the receipt of an NSERC Discovery Grant.  We also thank the anonymous referee for helpful comments.

This research made use of Astropy,\footnote{http://www.astropy.org} a community-developed core Python package for Astronomy \citep{Astropy2013,Astropy2018}. 
This research also made use of the computation resources provided by Westgrid (www.westgrid.ca) and Compute Canada (www.computecanada.ca).  

Funding for the Sloan Digital Sky Survey IV has been provided by the Alfred P. Sloan Foundation, the U.S. Department of Energy Office of Science, and the Participating Institutions. SDSS acknowledges support and resources from the Center for High-Performance Computing at the University of Utah. The SDSS web site is www.sdss.org.  SDSS is managed by the Astrophysical Research Consortium for the Participating Institutions of the SDSS Collaboration including the Brazilian Participation Group, the Carnegie Institution for Science, Carnegie Mellon University, the Chilean Participation Group, the French Participation Group, Harvard-Smithsonian Center for Astrophysics, Instituto de Astrofísica de Canarias, The Johns Hopkins University, Kavli Institute for the Physics and Mathematics of the Universe (IPMU) / University of Tokyo, the Korean Participation Group, Lawrence Berkeley National Laboratory, Leibniz Institut für Astrophysik Potsdam (AIP), Max-Planck-Institut für Astronomie (MPIA Heidelberg), Max-Planck-Institut für Astrophysik (MPA Garching), Max-Planck-Institut für Extraterrestrische Physik (MPE), National Astronomical Observatories of China, New Mexico State University, New York University, University of Notre Dame, Observatório Nacional / MCTI, The Ohio State University, Pennsylvania State University, Shanghai Astronomical Observatory, United Kingdom Participation Group, Universidad Nacional Autónoma de México, University of Arizona, University of Colorado Boulder, University of Oxford, University of Portsmouth, University of Utah, University of Virginia, University of Washington, University of Wisconsin, Vanderbilt University, and Yale University.

\bibliographystyle{mnras}
\bibliography{library}

\input{appendix.tex}

\label{lastpage}

\end{document}

%% file: appendix.tex
 \renewcommand{\theequation}{A\arabic{equation}}
  \renewcommand{\thesection}{A\arabic{section}}
  \renewcommand{\thefigure}{A\arabic{figure}}
    \setcounter{equation}{0}    \setcounter{section}{0}    \setcounter{figure}{0}  
  \section*{APPENDIX A: The Accuracy of Stellar Ages and Metallicities from pPXF}
  \label{appendix}

In order to test whether pPXF (\citealp{Cappellari2017}) produces reliable estimates for stellar age and metallicity, we produced a suite of synthetic test spectra with known ages and metallicities and tested to see how well pPXF could reproduce them given the typical noise spectrum and spectral resolution of MaNGA.   We constructed a suite of 40000 spectra from combinations of ``delayed $\tau$-model'' star-formation histories (\eg, \citealp{LopezFernandez2018}) of the form:
\begin{equation}
 \dot{M}_{*}(t) \propto (t-t_{\rm burst}) e^{-(t-t_{\rm burst}) /\tau}
\end{equation}
where $t_{\rm burst}$ is the beginning of the burst.  The 40000 spectra are random realizations with random values of several free parameters listed in \tab{modelfreeparams} and described here.

Our goal was to produce spectra with realistically evolving populations which included evolution in metallicity.  Therefore, we use the PEGASE 2 stellar population synthesis code of \cite{Fioc1999} to predict the evolution of the metallicities of delayed $\tau$-model SFHs, such that the stars formed in the next time step adopt the metallicity of the ISM which was enriched by the stars in the previous time step. 
Thus our delayed $\tau$ models include the initial metallicity $Z_o$ as a free parameter, as well as $\tau$ and $t_{\rm burst}$.  These were given random values in the ranges specified in \tab{modelfreeparams}.

Each model spectrum is randomly composed of 1, 2 or 3 of these delayed $\tau$-models, each with differing values of the free parameters.  These $\tau$-models are added with random mass fractions such that the later bursts are between factors of 0.1-2 times the mass fraction of the first burst.  In addition, we added the option of whether or not to quench the SFHs.  If the SFH was to be quenched, every SFH value after the time of quenching $t_{\rm q}$ is set to 0.  Thus, bursts that begin after $t_{\rm q}$ do not contribute to the SFH.  We require that $t_{\rm q}$ be at least 1 Gyr after the start of the earliest burst.  

\begin{table}
 \begin{center}
 \caption{\small Free parameters used in model spectra}
 \label{modelfreeparams}
   \begin{tabular}{ cp{2cm}p{3.5cm} }
 \hline\hline
\colh{Parameter} & \colh{Values} & \colh{Description} \\
 \hline
$\tau$ & 0.1-5 Gyr  & Scale length of the burst\\
$t_{\rm burst}$ & 0-12.6 Gyr & Start of the burst\\
$Z_o$ & 0.003-0.025 & Initial metallicity \\
$n_{\rm bursts}$ & 1-3 & Number of bursts in the spectrum\\
quench? & True, False & Flag of whether or not to quench\\
$f_{\rm burst}$ & 0.1-2 & Burst mass as a fraction of the mass in the first burst \\
$t_{\rm q}$ & min($t_{\rm burst}$)+1 to 13.6 Gyr & Time of quenching\\
\hline
\end{tabular}
  \end{center}
 \end{table}

Our model SFHs, evolving in both SFR and metallicity, can be directly translated into weights on the E-MILES age-metallicity grid of SSPs.  These weights can then be used to combine the spectra of the SSPs into a single spectrum for the population corresponding to a mass-weighted age and metallicity.  

In order to make the spectra realistic, for each simulated spectrum we randomly drew from the MaNGA baxel sample and applied the following modifications to the simulated spectrum:
\begin{enumerate}
\item We broadened the spectrum with the spectral resolution provided in the baxel's parent data cube (the ``\texttt{SpecRes}'' extension).
\item We also broadened the spectrum with the baxel's stellar velocity dispersion as measured by pPXF.
\item We reddened the spectrum with a random E(B-V) value between 0 and 1 using the attenuation curve of \cite{Cardelli1989}.
\item We then redshifted the spectrum to the redshift of the stellar component of the chosen baxel.  
\item We added the best-fitting gas spectrum of the baxel obtained by pPXF, which has its own velocity dispersion, reddening and redshift (which is usually similar to that of the stellar component, but not required to be identical).
\item We then applied a foreground extinction using the E(B-V) at the sky position of the baxel and the attenuation curve of \cite{Fitzpatrick1999}.
\item Lastly, we took the residuals of the best-fitting total spectrum from pPXF and added them to the simulated spectrum.  The noise spectrum of the baxel is also adopted as the noise spectrum of the simulated spectrum.
\end{enumerate}

We ran pPXF on the these test spectra and compared the measured mass-weighted log stellar ages and $i$-band $\Ms/L_i$ against their input values.  For fits with $\chi^2 < 3$ (96\% of our MaNGA baxels), the scatter between input and output log age and log $\Ms/L_i$ is 0.27 dex and 0.14 dex respectively.  However the scatter is heavily weighted by a number of outliers that are catastrophic failures.  We have found that many of these are removed with the following cuts on the measurements, which we can then apply on the MaNGA baxels: 
\begin{itemize}
\item Mass-weighted log age/yr $>$ 8.5 ($> 99$\% of baxels);
\item Light-weighted log age/yr $>$ 8.0 ($99$\% of baxels);
\item The difference between the mass-weighted and light-weighted age, log age$_{\rm MW}$ - log age$_{\rm LW}$ $<$ 1.1 (97\% of baxels);
\item Mass-weighted [M/H] $<$ 0.255 (99\% of our baxels); this cut removes the measurements that are saturated at the very highest metallicities (0.26);
\item $i$-band mass-to-light ratio, log $\Ms/L_i$ $>$ -0.3 (95\% of our baxels).
\item $\sntot > 10$ (95\% of baxels)
\end{itemize}

\begin{figure*}
\centerline{\includegraphics[width=0.85\linewidth]{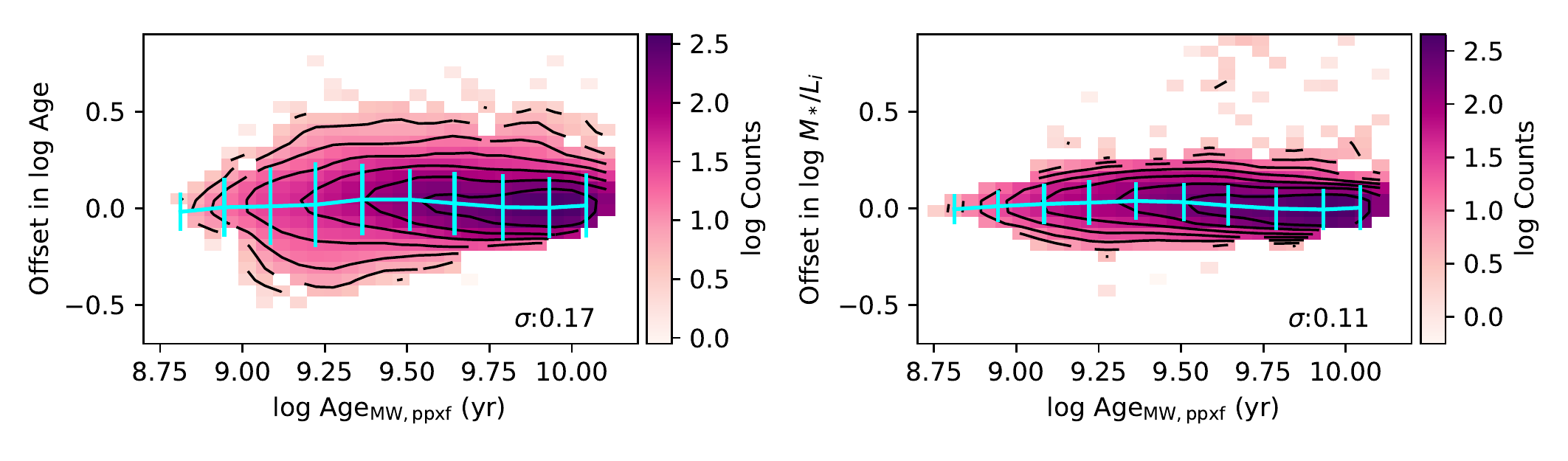}}
\caption{\small The systematic offset between input and output mass-weighted stellar age (left) and $\MLi$ as a function of output age.  The blue curve and error bars show median and 1-$\sigma$ standard deviation of the offsets.  The standard deviation of the offsets for the whole test set is shown in the bottom right corner. } 
\label{error_vs_lage}
\end{figure*}

\begin{figure*}
\centerline{\includegraphics[width=0.8\linewidth]{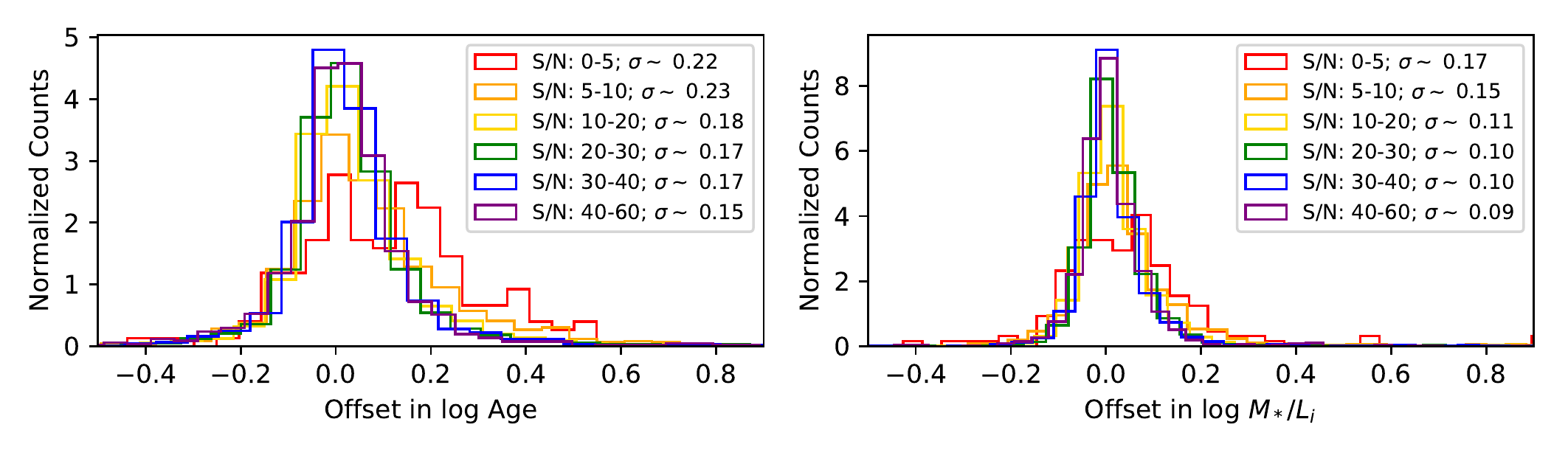}}
\caption{\small The distributions of offset between input and output mass-weighted stellar age (left) and $\MLi$ in bins of S/N.  The distributions are well-behaved for S/N $>$ 10.} 
\label{error_histograms}
\end{figure*}

With these cuts, the scatter in log age and log $\Ms/L_i$ is reduced to 0.17 dex and 0.11 dex respectively.  We show the distributions of the error in log age and log $\Ms/L_i$ in \twofigs{error_vs_lage}{error_vs_lML_i} as a function of measured log age, $\Ms/L_i$ and S/N.  The scatter does increase slightly for younger ages and lower S/N, but the offsets hover around 0 with very little dependence on any of these quantities.  We also show the distributions of error in log age and log $\Ms/L_i$ in bins of S/N in \fig{error_histograms}.  For S/N $<$ 10, the distributions of the offsets are skewed toward older ages and higher $\MLi$.  Thus, we limit our analyses to use baxels with $\sntot > 10$.  
Lastly we note that the presence of dust and/or emission lines had very little effect on the quality of the fits.

\begin{figure*}
\centerline{\includegraphics[width=0.85\linewidth]{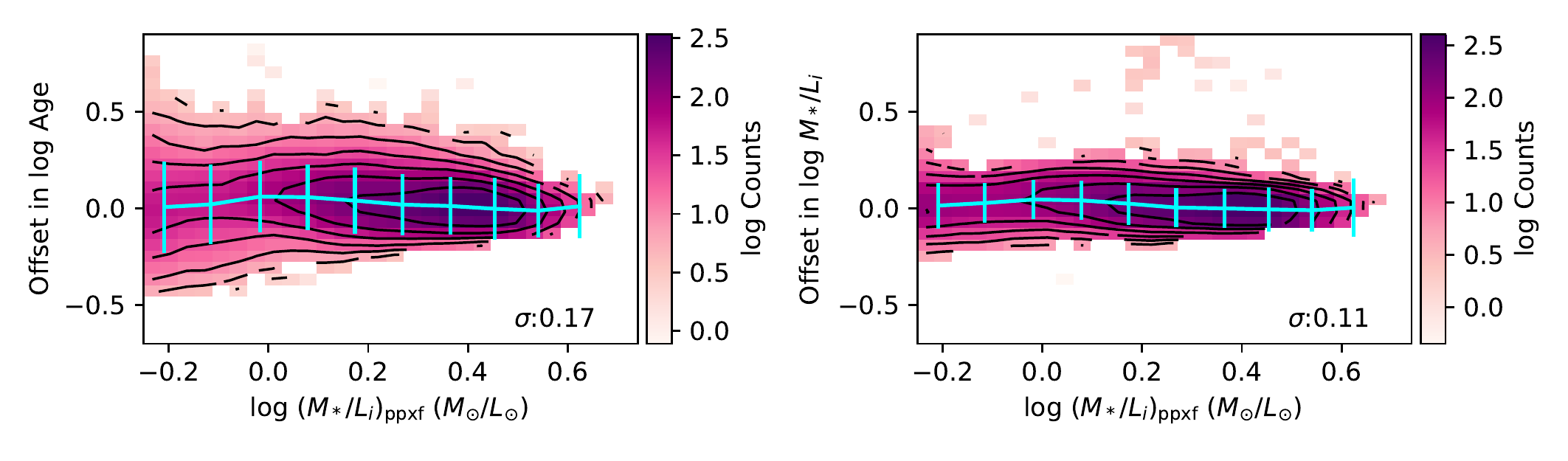}}
\caption{\small  The systematic offset between input and output mass-weighted stellar age (left) and $\MLi$ as a function of output $\MLi$.  The blue curve and error bars show median and 1-$\sigma$ standard deviation of the offsets.  The standard deviation of the offsets for the whole test set is shown in the bottom right corner.  } 
\label{error_vs_lML_i}
\end{figure*}

\section*{APPENDIX B: Profiles of $\dlage$, $\dssfr$ and $\doh$ for Individual Galaxies}
\label{appendixB}

\begin{figure*}
\includegraphics[width=0.8\linewidth]{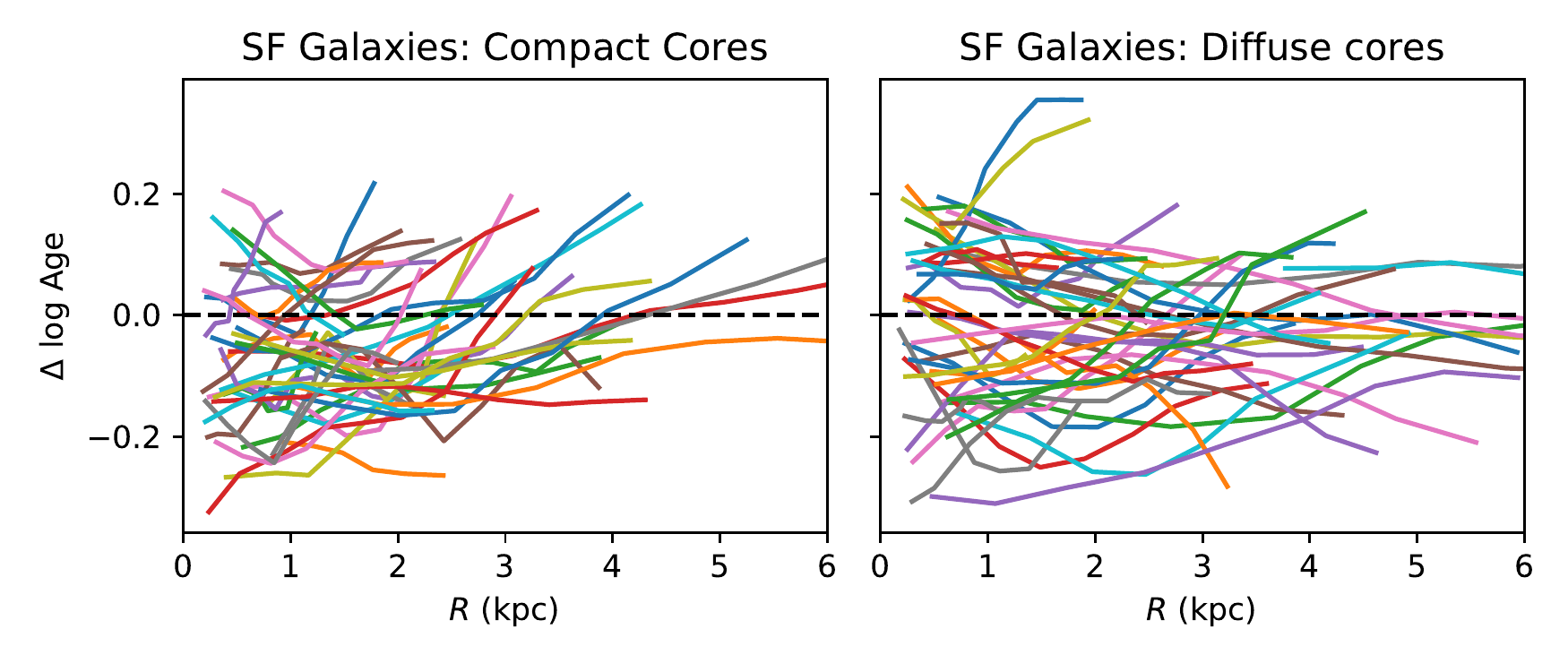}
\caption{\small The median profiles of $\dlage$ for 50 random star-forming galaxies with compact cores (left) and 50 random galaxies with diffuse cores (right).  The profiles are coloured randomly to aid in identifying individual profiles.} 
\label{ridgeline_age_indiv}
\end{figure*}

\begin{figure*}
\includegraphics[width=0.8\linewidth]{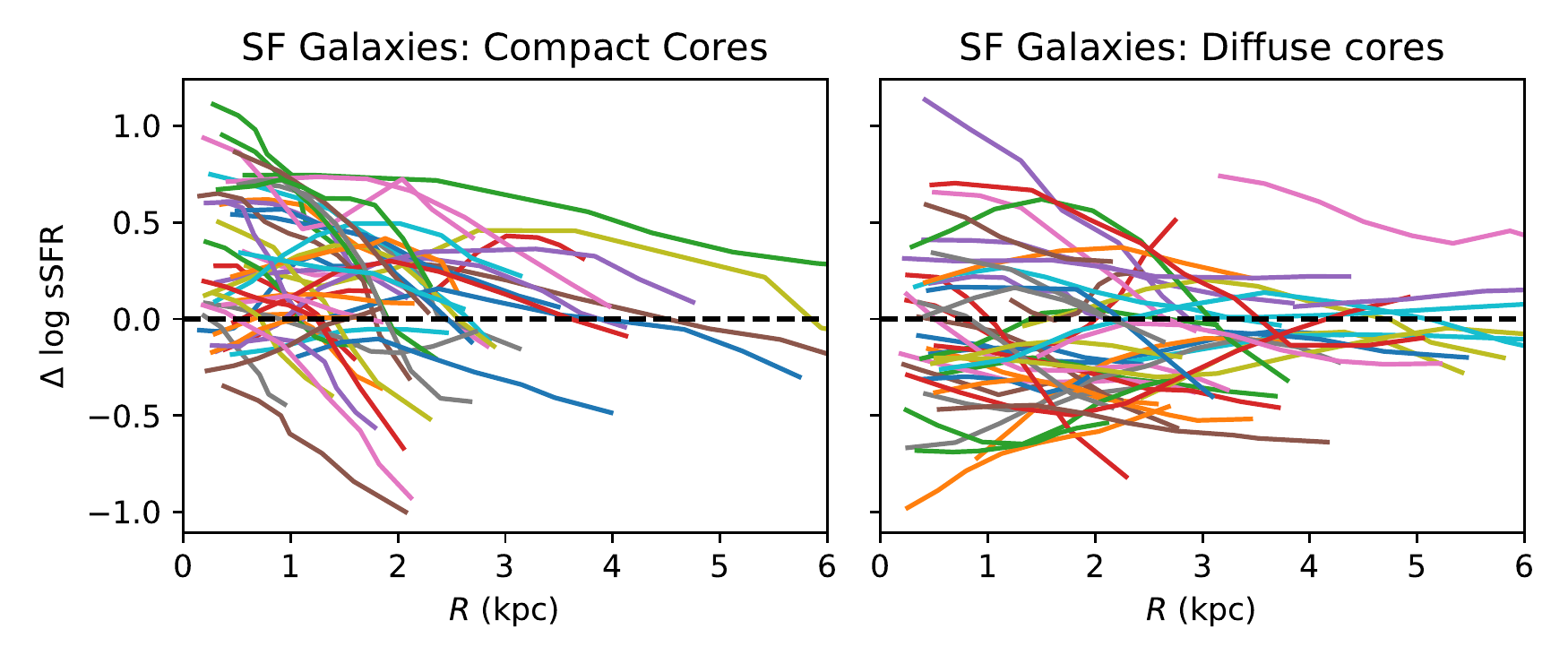}
\caption{\small The median profiles of $\dssfr$ for 50 random star-forming galaxies with compact cores (left) and 50 random galaxies with diffuse cores (right).  The profiles are coloured randomly to aid in identifying individual profiles.} 
\label{ridgeline_sf_indiv}
\end{figure*}

\begin{figure*}
\includegraphics[width=0.8\linewidth]{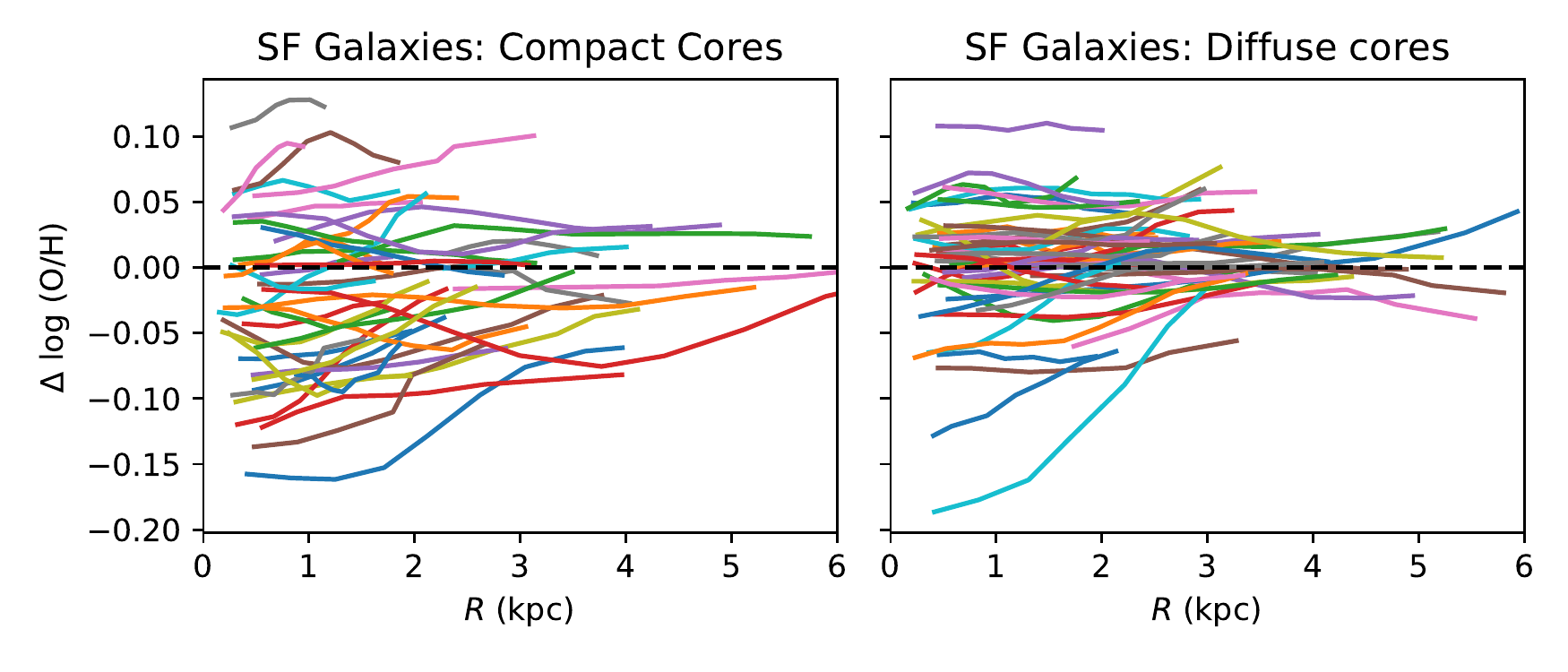}
\caption{\small The median profiles of $\doh$ for 50 random star-forming galaxies with compact cores (left) and 50 random galaxies with diffuse cores (right).  The profiles are coloured randomly to aid in identifying individual profiles.} 
\label{ridgeline_gasmet_indiv}
\end{figure*}

\begin{figure*}
\includegraphics[width=0.8\linewidth]{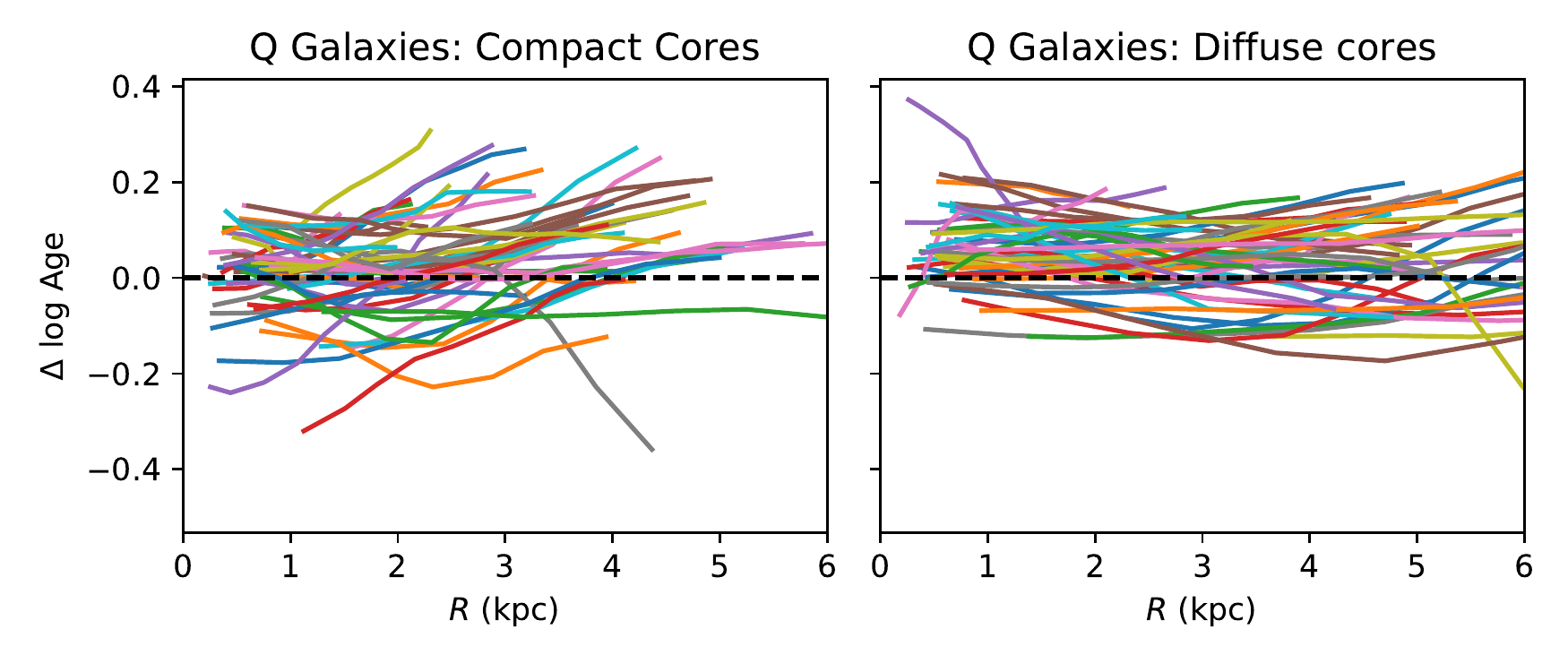}
\caption{\small The median profiles of $\dlage$ for 50 random quiescent galaxies with compact cores (left) and 50 random galaxies with diffuse cores (right).  The profiles are coloured randomly to aid in identifying individual profiles.} 
\label{ridgeline_age_q_indiv}
\end{figure*}

\begin{figure*}
\includegraphics[width=0.8\linewidth]{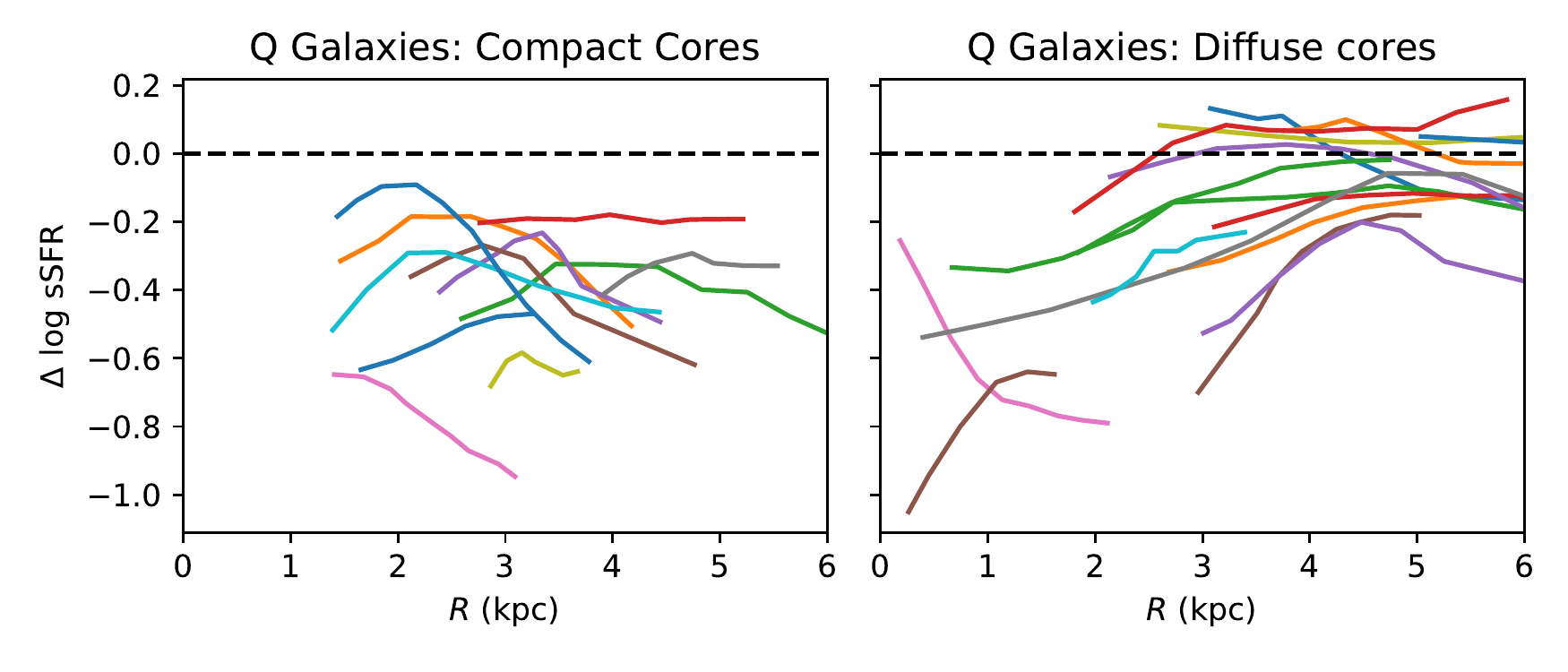}
\caption{\small The median profiles of $\dssfr$ for all quiescent galaxies with compact cores (left) and diffuse cores (right).  The profiles are coloured randomly to aid in identifying individual profiles.} 
\label{ridgeline_sf_q_indiv}
\end{figure*}

\begin{figure*}
\includegraphics[width=0.8\linewidth]{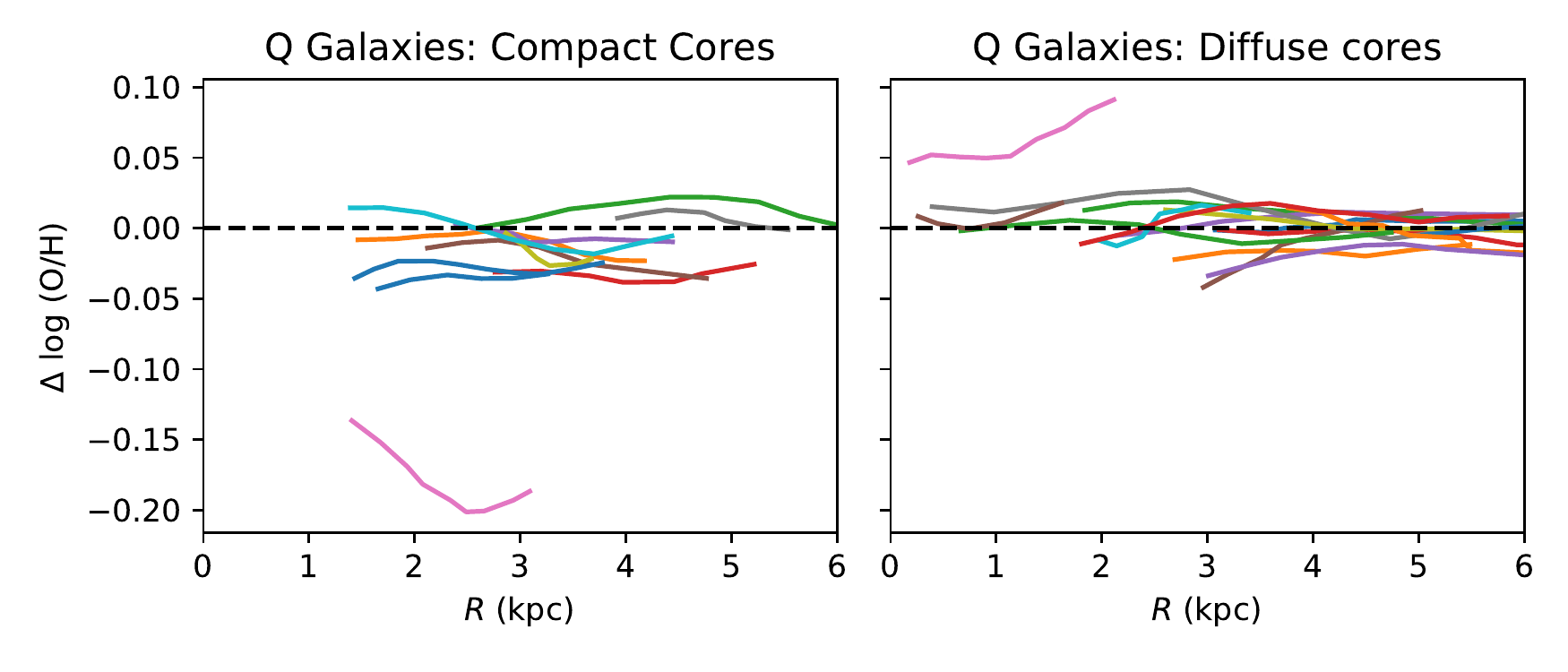}
\caption{\small The median profiles of $\doh$ for all quiescent galaxies with compact cores (left) and diffuse cores (right).  The profiles are coloured randomly to aid in identifying individual profiles.} 
\label{ridgeline_gasmet_q_indiv}
\end{figure*}

\secref{sfridgelinesec} and \secref{sfridgelinesec_q} presented median profiles of $\dlage$, $\dssfr$ and $\doh$ for star-forming or quiescent galaxies with compact or diffuse cores.  Here we present the profiles of the individual galaxies of the same populations.  These were computed similarly to median profiles.   Specifically, they are the median of all baxels in an individual galaxy within radial bins and then smoothed with a box filter kernel of 3 radial bins.  The bin ``centres'' are the median $R$ of all baxels within the radial bins.  Radial bins are discarded from the plotted profiles if the number of baxels in the bin is less than 3.  The individual profiles are shown in Figs. \ref{ridgeline_age_indiv}-\ref{ridgeline_gasmet_q_indiv}.  These are grouped into the same populations as those shown in Figs. \ref{ridgeline_age}-\ref{ridgeline_gasmet_q}.  The plots are unclear when too many profiles are shown.  Therefore, if any panel in Figs. \ref{ridgeline_age_indiv}-\ref{ridgeline_gasmet_q_indiv} contained more than 50 galaxies, we plotted the profiles of 50 galaxies chosen at random.  Figs. \ref{ridgeline_sf_q_indiv} and \ref{ridgeline_gasmet_q_indiv} contain fewer than 50 galaxies, so all of them are plotted.  The individual profiles are coloured randomly in order to aid the eye in separating them from each other.

